\documentclass[superscriptaddress,english,floatfix,twocolumn,10pt]{revtex4-1}
\usepackage[T1]{fontenc}
\usepackage{amsmath}
\usepackage{textcomp}
\usepackage[latin1]{inputenc}
\usepackage{subfigure}
\usepackage{graphicx}
\usepackage{blindtext, rotating}
\usepackage{mathtools}
\usepackage{blindtext}
\usepackage{gensymb}
\usepackage{enumitem}
\usepackage{xcolor}
\usepackage{babel}
\usepackage{hyperref}
\usepackage{braket}
\usepackage{natbib}
\usepackage{subfigure}  
\makeatother
\begin{document}

\title{Effect of spin-orbit coupling in  one-dimensional quasicrystals with  power-law hopping}
\date{\today}
\author{Deepak Kumar Sahu}
\email{dksphysics90@gmail.com}
\affiliation{Department of Physics and Astronomy, National Institute of Technology, Rourkela, Odisha, India}
\author{Sanjoy Datta} 
\email{dattas@nitrkl.ac.in}
\affiliation{Department of Physics and Astronomy, National Institute of Technology, Rourkela, Odisha, India}
\affiliation{Center for Nanomaterials, National Institute of Technology, Rourkela, Odisha, India}

\begin{abstract}
In the one-dimensional quasiperiodic Aubry-Andr\'{e}-Harper Hamiltonian with nearest-neighbor hopping, all single-particle eigenstates undergo a phase transition from ergodic to localized states at a critical disorder strength $W_c/t = 2.0$. There is no mobility edge in this system. However, in the presence of power-law hopping having the form $1/r^a$, beyond a critical disorder strength mobility edge appears for $a > 1$, while, for $0< a\leq 1$, a multifractal edge separates the extended and the multifractal states. In both these limits, depending on the strength of the disorder, lowest $\beta^s L$ states are delocalized. We have found that, in the presence of the spin-orbit coupling, the critical disorder strength is always larger irrespective of the value of the parameter $a$. Furthermore, we demonstrate that for $a\leq 1$, in the presence of spin-orbit coupling, there exists multiple multifractal edges, and the energy spectrum splits up into alternative bands of delocalized and multifractal states. Moreover, the location of the multifractal edges are generally given by the fraction $(\beta^s \pm \beta^m)$. The qualitative behavior of the energy spectrum remains unaffected for $a > 1$. However, in contrast to the previously reported results, we find that in this limit, similar to the other case, multiple mobility edges can exist with or without the spin-orbit coupling.  
\end{abstract}
\maketitle

\section{Introduction}
Understanding the nature of the eignestates in disordered systems has continued to attract the attention since the publication of the seminal article of P. W. Anderson \cite{Anderson}. It was argued that in three dimensions, beyond a critical disorder strength all the single particle electronic eigenstates fall of exponentially from the point of their maximum amplitude, while for weak disorder a \textit{mobility edge} separates the delocalized and localized states. Furthermore, in one and two dimensions all the electronic states are localized even for arbitrarily weak disorder \cite{Mott-Twose, Borland, Berezinskii}. However, the existence of delocalization-localization (DL) transition has been predicted in two-dimensional system with random disorders in the presence of spin-orbit coupling \cite{Hikami}, and there are numerical evidences in support of this assertion \cite{Evangelou,Asada-2D-SOC-PRL02,Su-Wang}. In one-dimension, however, there is still no evidence of the DL transition, 
in systems with random disorder. 

On the other hand, quasiperiodic lattices are paradigmatic examples of systems which hosts all the richness of randomly disordered systems, 
even in one-dimension. The minimal Hamiltonian describing such systems is the Aubry-Andr\'{e}-Harper (AAH) model \cite{Harper, Azbel, Aubry-Andre} having nearest-neighbor hopping and a cosine potential whose periodicity is modulated by the inverse golden ratio. AAH Hamiltonian shows a DL transition at a critical strength of the potential. All the energy eigenstates are extended below the critical point, while all of them are localized above it. In contrast to the lattice with random disorders, there is no mobility edge in the energy spectrum at any value of the disorder strength in the original AAH Hamiltonian.

Recently, however, it has been demonstrated \cite{Deng} that mobility edge can exist in the one-dimensional (1D) quasiperiodic lattice if the hopping amplitude $t$ is replaced by a power-law hopping of the form $t/{r^a}$, where $r$ is the distance between two lattice sites $i$ and $j$ $(i \neq j)$. It is evident that this generalized Aubry-Andr\'{e} (GAA) Hamiltonian converges to the minimal AAH Hamiltonian in the limit $a >>1.$ The spectrum shows much richer structure when $``a''$ is not too large compare to unity. It is observed that for $a > 1 $ and $a \leq 1$ there exists a critical disorder strength, below which all the states delocalized. Interestingly, above the critical disorder strength the energy spectrum shows a mobility edge when $a>1$, while a \textit{multifractal edge} appears for $a \leq 1$ which separates the extended and the multifractal states.

Since spin-orbit coupling has dramatic effect on the randomly disordered systems, it is natural to wonder what happens to the energy spectrum and to the critical properties of the GAA Hamiltonian in the presence of spin-orbit coupling. To address these questions, in this article, we have studied the 1D GAA Hamiltonian in the presence of spin-orbit coupling of Rashba type. The Rashba Hamiltonian consists of a spin preserving hopping and a spin-flip hopping. It has recently been reported \cite{dsahu} that the qualitative structure of the energy spectrum remains intact for the minimal AAH Hamiltonian ($a >>1$ limit of the GAA Hamiltonian) in 
the presence of the Rashba spin-orbit (RSO) coupling, while the critical point moves to a higher strength of the disorder potential. In this article, however, we demonstrate that in the presence of the RSO coupling, the energy spectrum of the GAA Hamiltonian shows distinctly different behaviour compared to the original GAA Hamiltonian, especially when $a \leq 1$. We have found that, in the presence of the RSO coupling, beyond the critical disorder strength the energy spectrum becomes fragmented and multiple multifractal edges appear instead of a single multifractal edge observed in the pure GAA Hamiltonian. On the other hand, for $a>1$, we have found that above the critical point there are certain stretches of disorder strengths in which a single mobility edge exists in the pure GAA Hamiltonian. These windows of single mobility edge gradually disappear with an increase in the RSO coupling strength. Furthermore, in all the limits the critical disorder strength increases in the presence of the RSO coupling. 

\begin{figure*}
\vspace{-0.5cm}
\centering
\includegraphics[width=0.35\textwidth,height=0.33\textwidth]{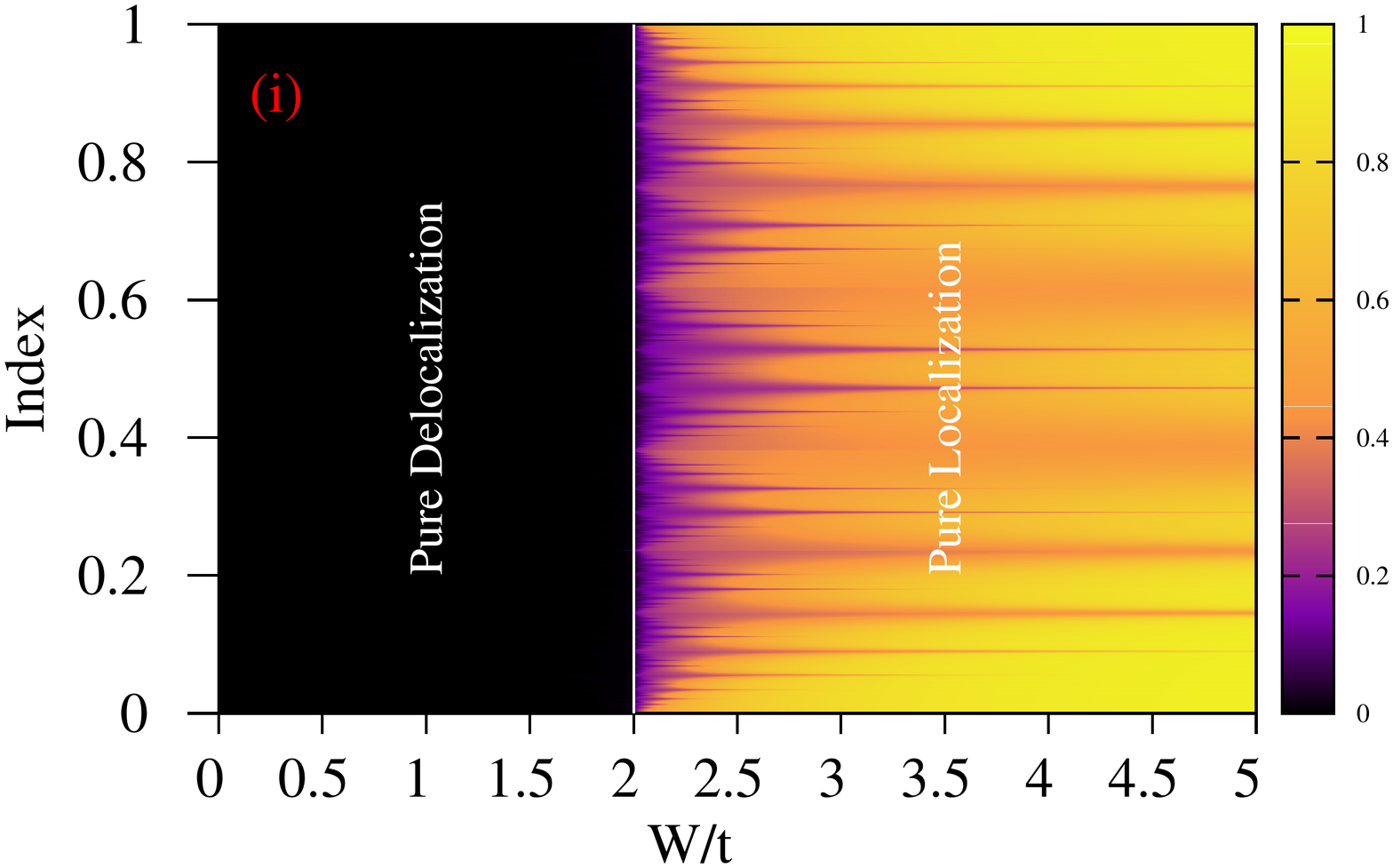}\hspace{-0.6cm}
\includegraphics[width=0.35\textwidth,height=0.30\textwidth]{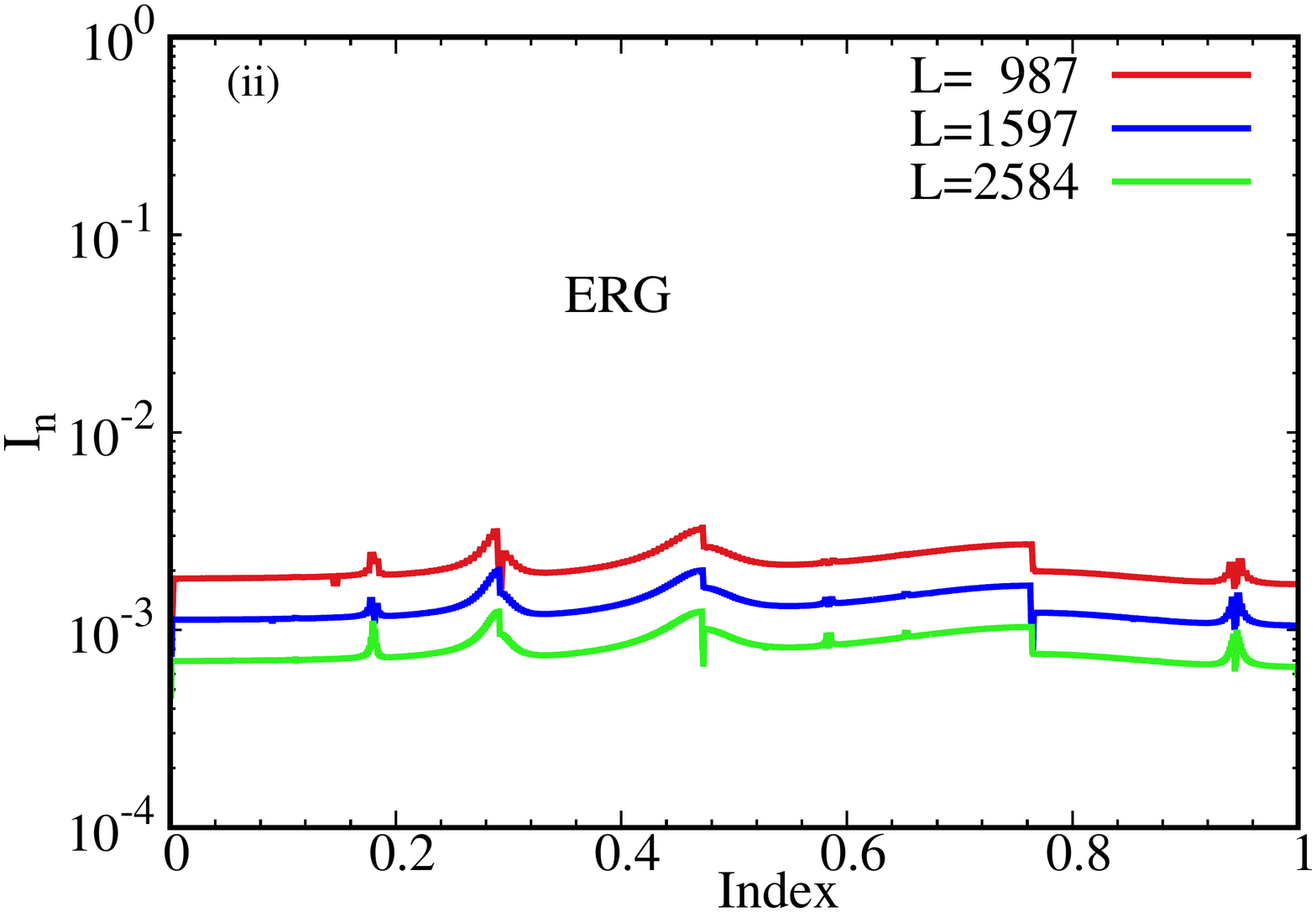}\hspace{-0.6cm}
\includegraphics[width=0.35\textwidth,height=0.30\textwidth]{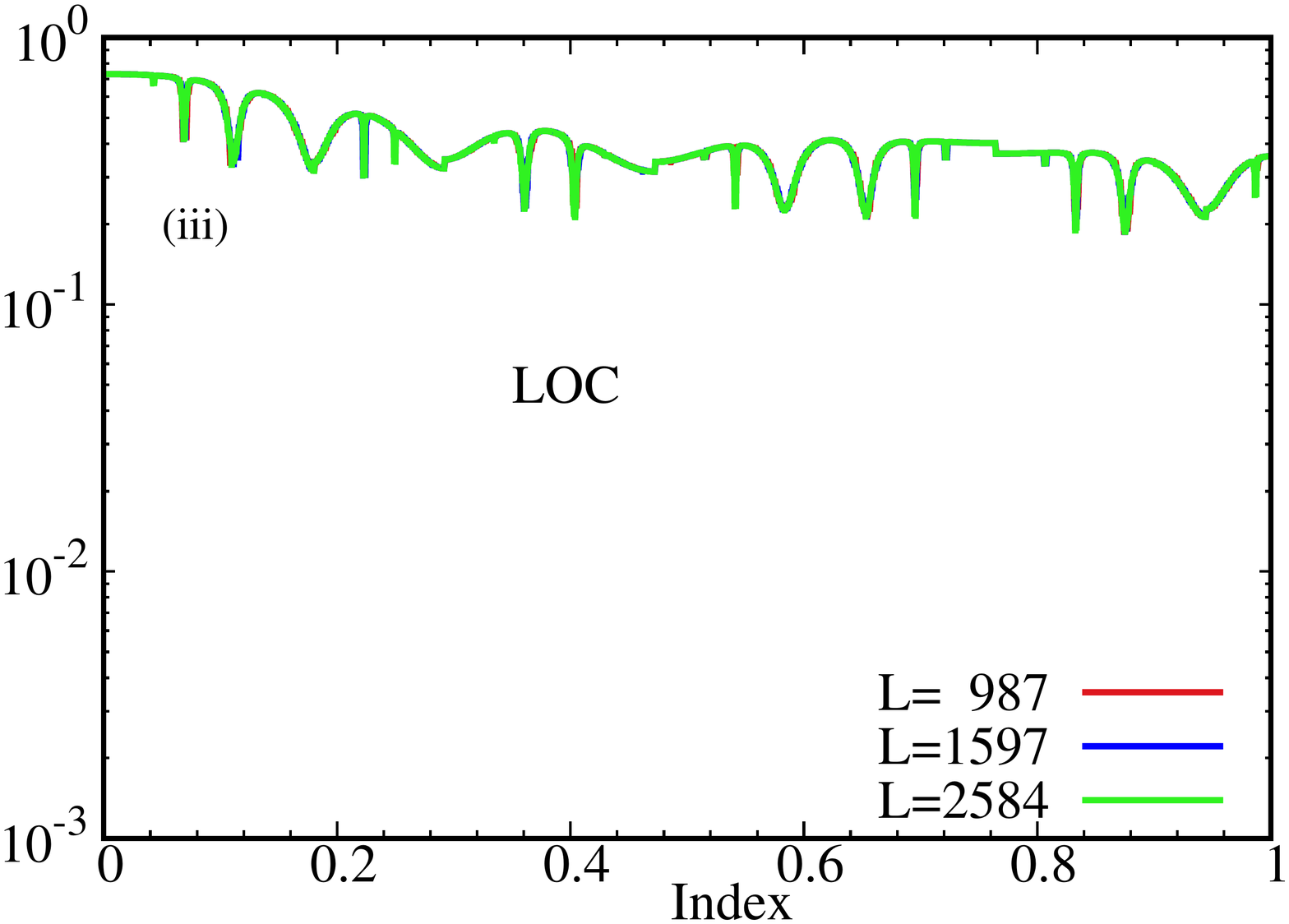}
\vspace{-0.5cm}
\caption{(i) Projection of IPR as a function of $W/t$ and eigen-energies $(index)$ for the GAA Hamiltonian ($L=1597$) in the limit $a\gg1$. (ii) and (iii): IPR $(I_n)$ vs eigen-energies $(index)$ at $W/t= 1.0$ and $W/t=2.5$ respectively.}
\label{Fig:AA-IPR}
\end{figure*}
\begin{figure*}
\vspace{-1.0cm}
\centering
\includegraphics[width=0.35\textwidth,height=0.33\textwidth]{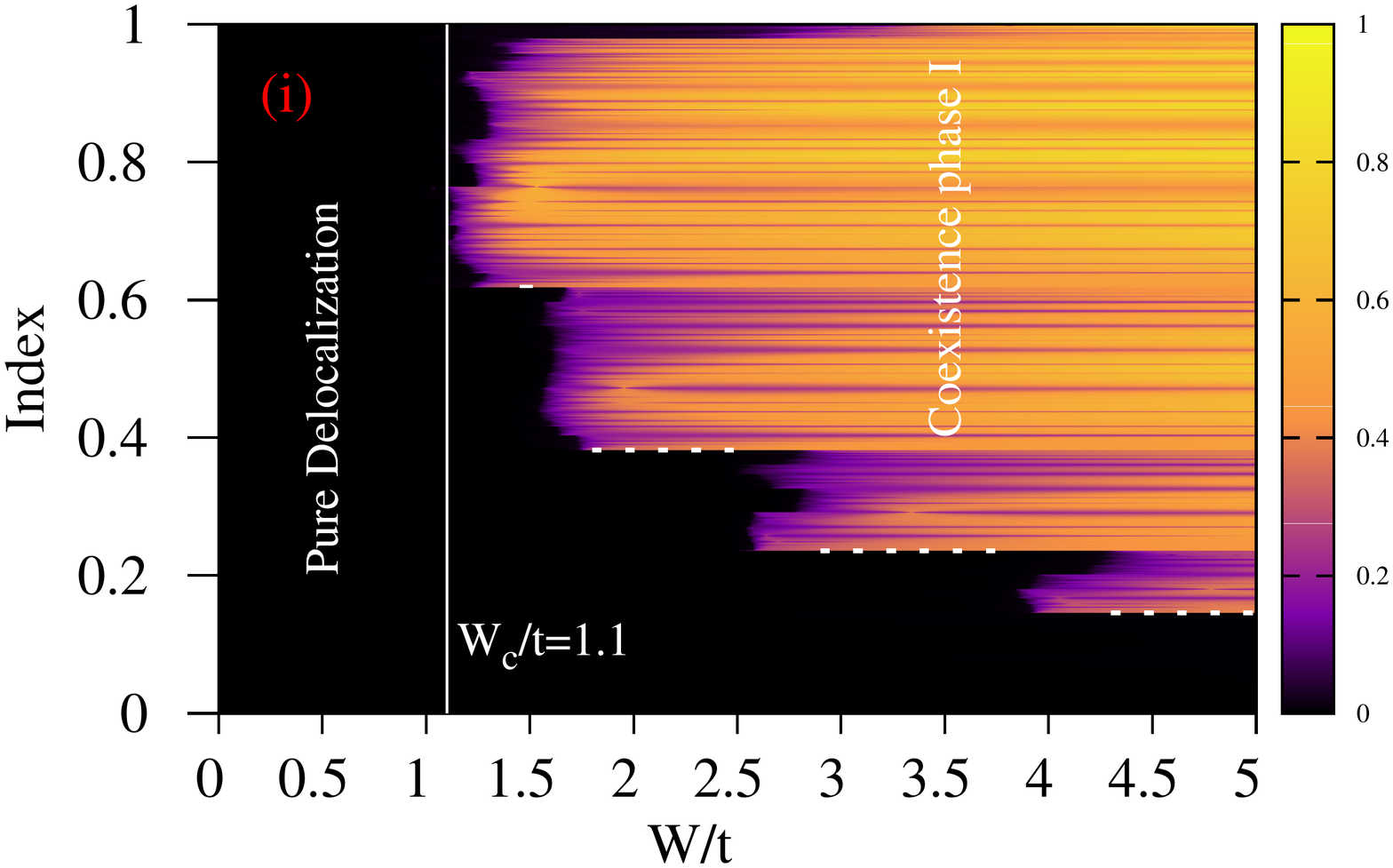}\hspace{-0.6cm}
\includegraphics[width=0.35\textwidth,height=0.30\textwidth]{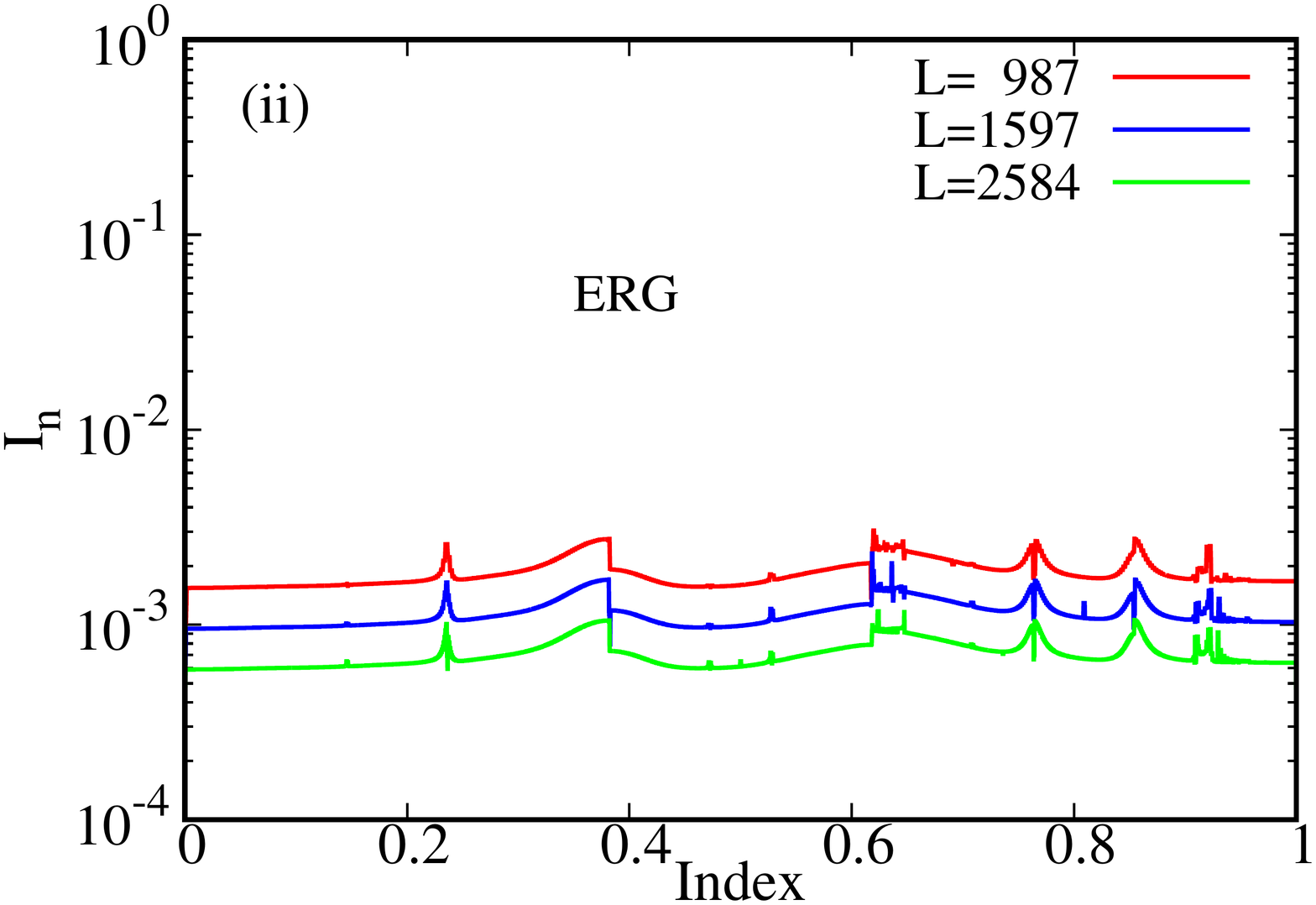}\hspace{-0.6cm}
\includegraphics[width=0.35\textwidth,height=0.30\textwidth]{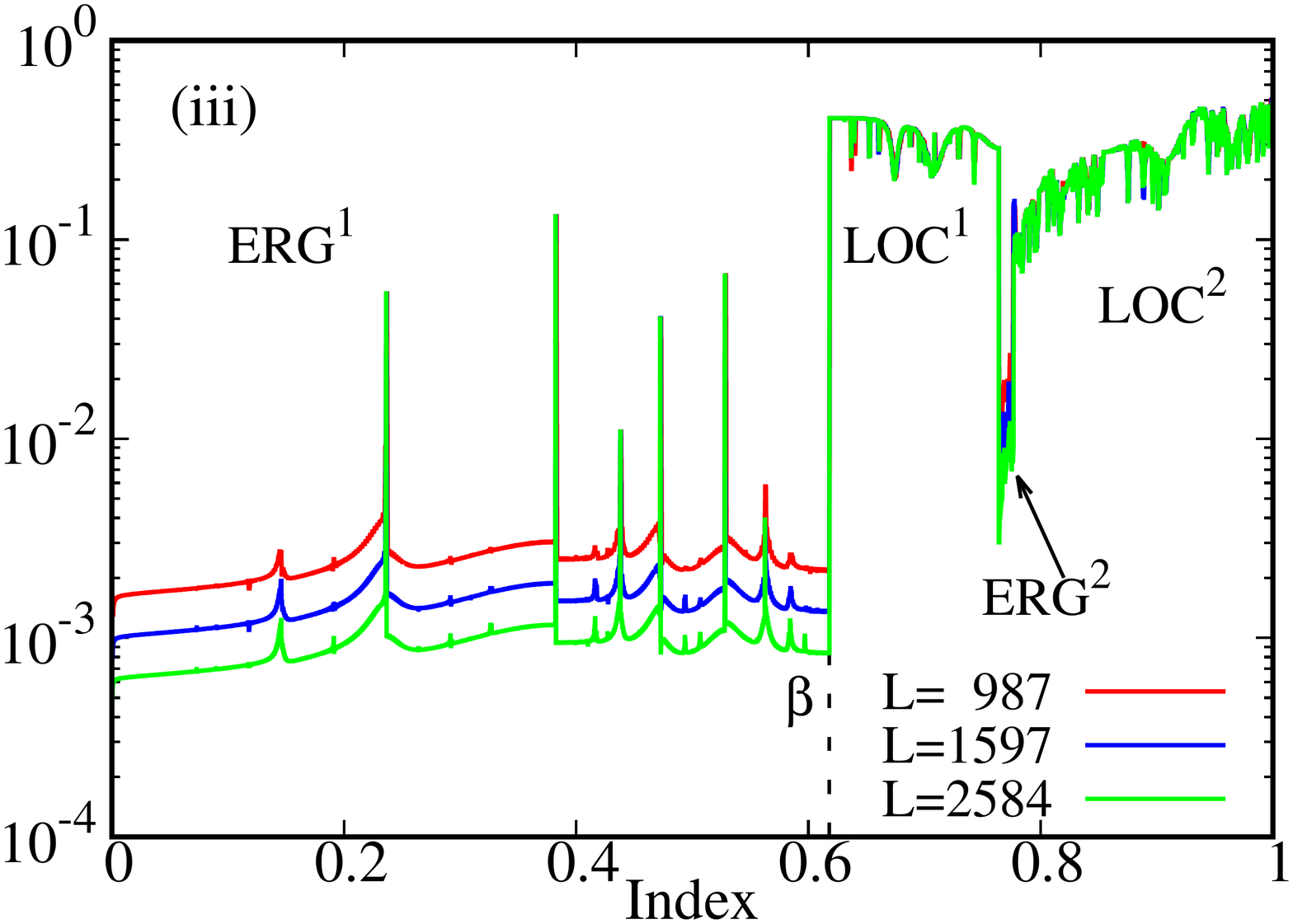}
\vspace{-0.5cm}
\caption{(i) Projection of IPR as a function of $W/t$ and eigen-energies $(index)$ in the limit of the short-range ($a = 1.5$) GAA Hamiltonian ($L=1597$). Dotted lines show the presence of single mobility edge regions. (ii) and (iii): IPR $(I_n)$ vs eigen-energies $(index)$ at $W/t= 0.5$ and $W/t=1.2$ respectively.}
\label{Fig:IPR-a1p5}
\end{figure*}
\begin{figure*}
\vspace{-1.0cm}
\centering
\includegraphics[width=0.35\textwidth,height=0.33\textwidth]{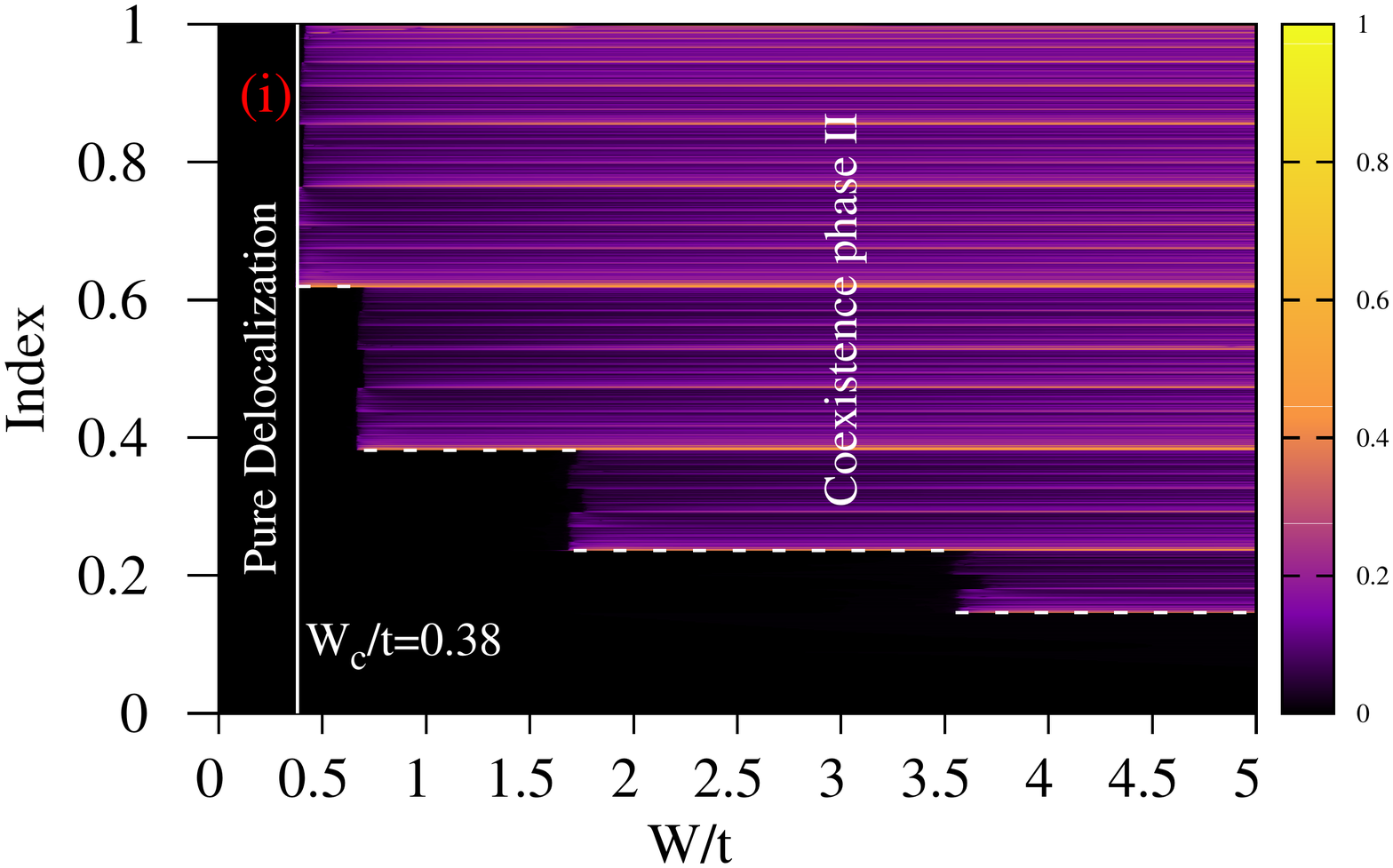}\hspace{-0.6cm}
\includegraphics[width=0.35\textwidth,height=0.30\textwidth]{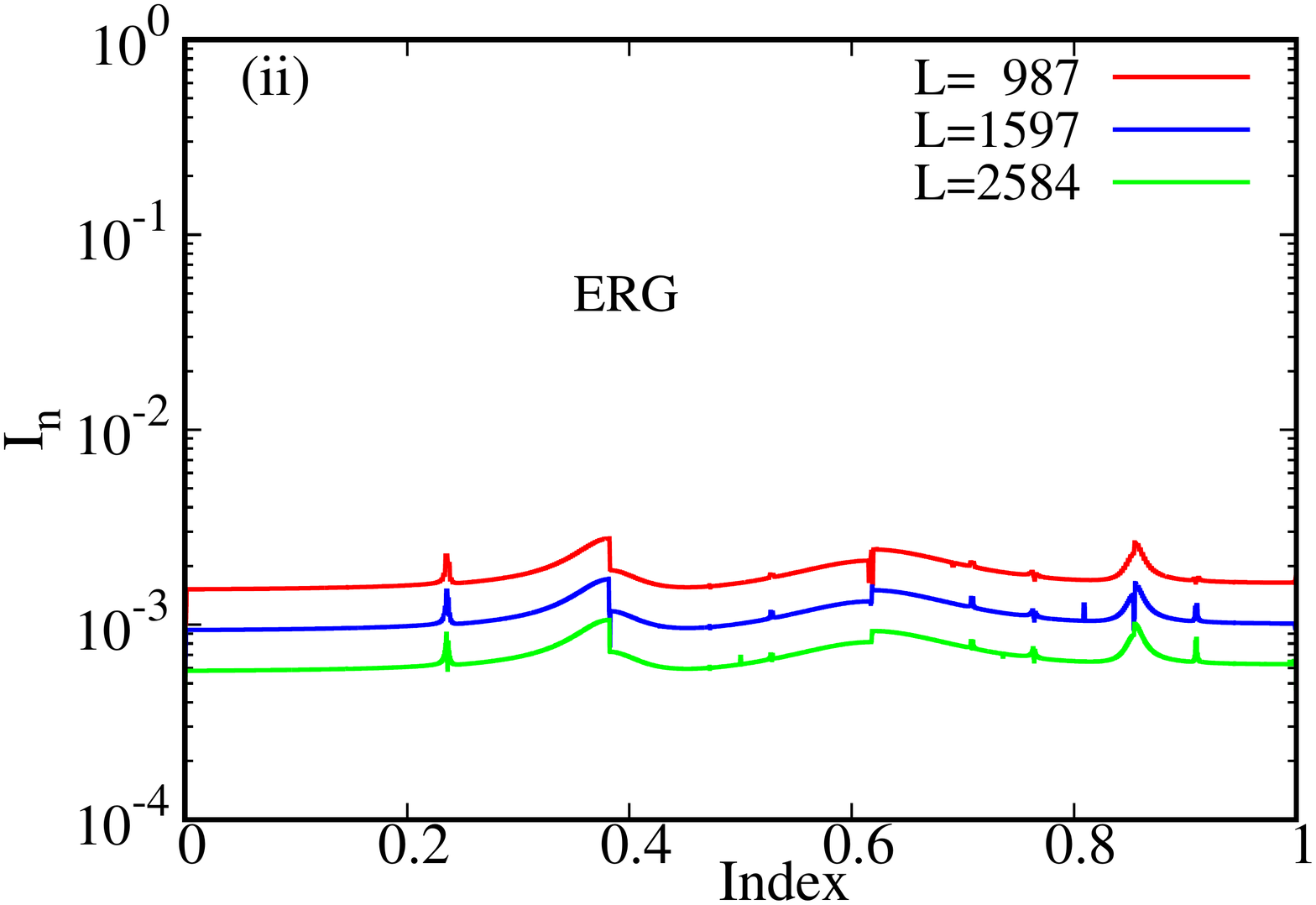}\hspace{-0.6cm}
\includegraphics[width=0.35\textwidth,height=0.30\textwidth]{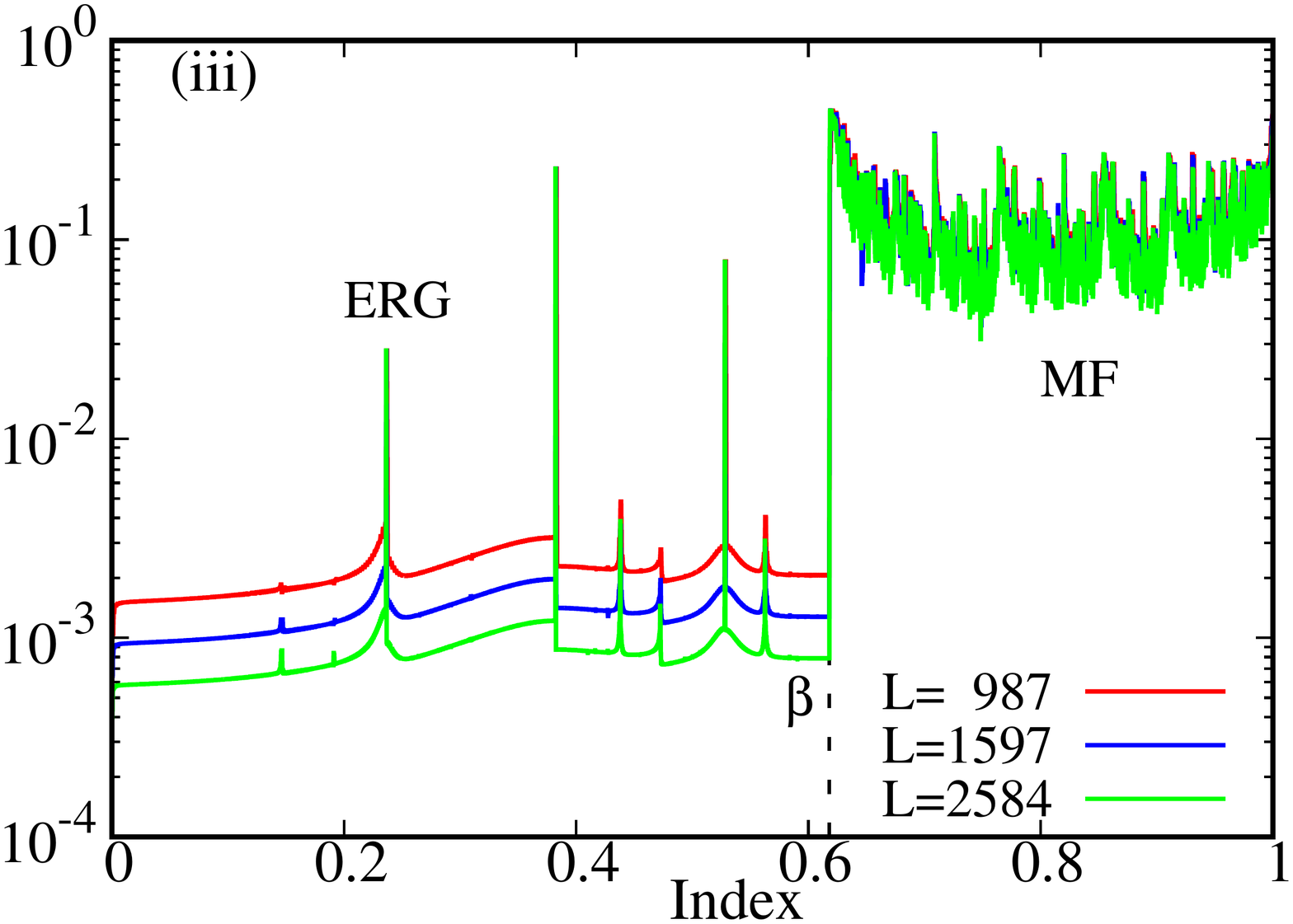}
\vspace{-0.5cm}
\caption{(i) Projection of IPR as function of $W/t$ and eigen-energies $(index)$ in the limit of long-range ($a=0.5$) GAA Hamiltonian ($L=1597$). 
Dotted lines show the presence of single multifractal edge regions. (ii) and (iii): IPR $(I_n)$ vs eigen-energies $(index)$ at $W/t= 0.2$ and $W/t=0.5$ respectively.}
\label{Fig:IPR-a0p5}
\end{figure*}
This article is  organized as follows. In Sec.~\ref{Sec:Model-Hamiltonian}, we have discussed the model Hamiltonian, and different parameters are used in this ambiance. In Sec.~\ref{Sec:IPR}, we have shown the presence of mobility and multifractal edges with the help of the Inverse Participation Ratio (IPR). We have done the multifractal analysis of the RSO coupled GAA model to analyze the multifractal edges in Sec.~\ref{Sec:multifractal-spectrum}.

\section{Generalized Aubry-Andr\'{e} Hamiltonian With RSO coupling}
\label{Sec:Model-Hamiltonian}
The Hamiltonian considered in this work is given by,
\begin{equation}
H=H_G+H_R,
\label{Eq:eq1}
\end{equation}
where, $H_G$ is the GAA model considered in Ref.~\cite{Deng}, and it is given as,
\begin{multline}
H_G=-t\displaystyle\sum_{\langle i,j\rangle,\sigma}^{L} \frac{1}{\lvert i-j\rvert^a} (c^\dag_{{i},{\sigma}} c_{{j},{\sigma}}+ 
c^\dag_{{j},{\sigma}} c_{{i},{\sigma}}) +\\
W\displaystyle\sum_{{i=1},{\sigma}}^{L}cos(2\pi \beta i+\phi)c^\dag_{{i},{\sigma}} c_{{i},{\sigma}}.
\label{Eq:GAA}
\end{multline}

In the above Hamiltonian $H_G$, $t/\lvert i-j\rvert^a$ is the rate of hopping from site $i$ to site $j$ and $L$ is the size of the quasiperiodic lattice, $c^\dag_{{i},{\sigma}}$ and $c_{{i},{\sigma}}$ are the fermionic creation and annihilation operators respectively at the lattice site $i$ with the spin $\sigma (\uparrow / \downarrow)$.$W$ is the strength of the quasiperiodic potential, and $\phi$ is an arbitrary phase varying from $(0, 2\pi)$. Our conclusions are independent of the value of $\phi$. Henceforth, we set $\phi=0$ for all our calcuations. The quasiperiodicity is introduced via the irrational constant quantity $\beta$ which has been set to $(\sqrt{5}-1)/2$.  

The RSO Hamiltonian $H_R$ is given by \cite{Birkholz},
\begin{multline} 
H_R=-\alpha_z \displaystyle\sum_{{i=1},{\sigma},{\sigma'}}^{L-1} t(c^\dag_{i+1,\sigma}(i\sigma_y)_{\sigma,\sigma'}
 c_{i,\sigma'}+ h.c.)+ \\
 \alpha_y \displaystyle\sum_{{i=1},{\sigma},{\sigma'}}^{L-1} t(c^\dag_{i+1,\sigma}(i\sigma_z)_{\sigma,\sigma'}  c_{i,\sigma'}+ h.c.),
 \label{Eq:RSO}
\end{multline}
where $\sigma_y$ and $\sigma_z$ are Pauli's spin matrices in $y$- and $z$-direction respectively. $\alpha_y$ is a complex spin-conserving hopping 
due to the confinement in $y$-direction and $\alpha_z$ is a spin-flip hopping due to the confinement in $z$-direction. The RSO Hamiltonian
$H_R$ considered here, has been studied in the context of quantum nanowires \cite{Ando-PRB89, Janssen-Huckestein}. Also, recently, localization property of attractive fermions have been studied by taking only the spin-flip component $(\alpha_z)$ of the RSO Hamiltonian\cite{Minakuchi}.
In this article, we mainly focus on the two limits of the Hamiltonian: long-range $(0< a \leq 1)$ and short-range $(a>1)$ hoppings. In the limit $a\gg 1$, $H_G$ converges to the pure AAH Hamiltonian. The pure AAH Hamiltonian in the presence of the RSO coupling has been studied in Ref.~\cite{dsahu}.

\section{Inverse Participation Ratio}
\label{Sec:IPR}
The Inverse participation ratio (IPR) is an useful indicator to distinguish localized and delocalized states in the energy spectrum. It gives a
measure of the number of sites contributing to a given state\cite{Kinnon}. Also, it is very convenient to pinpoint the mobility-edge. If a mobility edge exists, IPR value is expected to jump from system size independent higher value to a lower value that scales inversely with the system size. For $n$-th quasiparticle eigenstate the usual definition of IPR \cite{ADMirlin} can be generalized as follows;
\begin{equation}
I_{n}(q) =\sum_{\sigma} \sum_{i=1}^{N} \left|\psi_{n,\sigma}(i) \right|^{2 q},~~q=2.  
\end{equation}
Here, the $n$th single quasiparticle eigenstate is given by $\ket{\Psi_{n}} = \sum_{\sigma} \sum_{i=1}^{N} \psi_{n,\sigma}(i) \ket{1,\sigma}_{i},$
where $\ket{1,\sigma}_{i} = \ket{0,0,\cdots,\sigma_i,\cdots,0,0}$ represents the localized basis state having one particle with spin $\sigma$ at site $i$, and $N$ is the size of the lattice. Since, this is a non-interacting problem, usual scaling properties with respect to the system size is also expected to hold for the quasiparticle states, i.e., for a perfectly extended metallic state $I_{n}(q) = 1/N$, while for a state localized at a single site $I_{n}(q) \simeq 1$. The IPR has similar qualitative features for the multifractal states as the localized states. However, the IPR values show large fluctuations compared to a localized state in this case. The distinct scaling behavior of IPR helps to identify delocalized states unambiguously. It can also be used for preliminary identification of localized/multifractal states. 

\subsection{GAA Hamiltonian without RSO coupling}
Before presenting our results for the GAA Hamiltonian in the presence of the RSO coupling, in this section, we first review the results of the pure GAA Hamiltonian and discuss some new observations that were not addressed earlier. In Fig.~\ref{Fig:AA-IPR}(i), we have presented the projection of IPR as a function of $W/t$ for the GAA Hamiltonian for the limit $a\gg1$. As mentioned earlier, the GAA Hamiltonian coincides with the pure AAH Hamiltonian in this limit. In the pure AAH Hamiltonian, the DL transition takes place at a critical disorder strength $W_c/t=2$. This is evident from the sharp change in the IPR values in Fig.~\ref{Fig:AA-IPR}(i). 
Below the critical disorder strength all the eigenstates are delocalized, while all of them are localized above the critical value. To verify the behavior of IPR, we have shown the corresponding results in Figs.~\ref{Fig:AA-IPR}(ii) and (iii) for $W=1.0$ (delocalized region) and $W=2.5$ (localized region) respectively.

In Figs.~\ref{Fig:IPR-a1p5} and \ref{Fig:IPR-a0p5}, we have shown the IPR results for the short-range $(a > 1)$ and long-range $(a \leq 1)$ 
limits of the GAA model, respectively. From Fig.~\ref{Fig:IPR-a1p5}(i) it is evident that in the short-range hopping limit \textbf{($a=1.5$)} there 
is a critical disorder strength (lower than that in the $a\gg1$ limit) below which all the eigenstates are delocalized. Interestingly, however, 
beyond this critical point a mobility edge appears in the energy spectrum which separates the localized and delocalized states. A closer 
look at Fig.~\ref{Fig:IPR-a1p5}(i) reveals that there are certain windows of disorder strength (marked with horizontal dotted lines) in which all the 
states are localized above the mobility edge. Interestingly, however, at the beginning of each of these windows delocalized states can exists after 
the appearance of the first mobility edge. From Fig.~\ref{Fig:IPR-a1p5}(iii) one can observe that for $W/t=1.2$ a small band of delocalized states appear in the spectrum after the appearance of the mobility edge at $\beta$. This observation of the delocalized states in the energy spectrum beyond the critical point is in contrast with the conclusion of the Ref.~\cite{Deng}. In Ref.~\cite{Deng}, it was pointed out that beyond the critical point, all the states are localized above the mobility edge, and only a certain fraction of the lowest energy eigenstates are delocalized. This is indeed true only within the above mentioned windows of disorder strength. Within these windows, all the states are localized beyond the mobility edge, and such mobility edges appear at $n/L=\beta$, next at 
$n/L=\beta^2$, and so on. Depending on the window of the disorder strength, the lowest $\beta^s L$ amount of states, where $s=1,2,3,\cdots$, remain delocalized. From Fig.~\ref{Fig:IPR-a1p5}(i), it can be seen that the location of the mobility edge remains unchanged at $\beta$ for a range of a disorder strength. Next, such mobility edge appears at $\beta^2$, then at $\beta^3$, and so on. In Figs.~\ref{Fig:IPR-a1p5}(ii) and (iii), we have plotted the IPR results for different system sizes below and above the critical disorder strength respectively. The existence of a mobility edge beyond the critical point is quite clear. 
\begin{figure*}[ht]
\centering
\includegraphics[width=0.35\textwidth,height=0.30\textwidth]{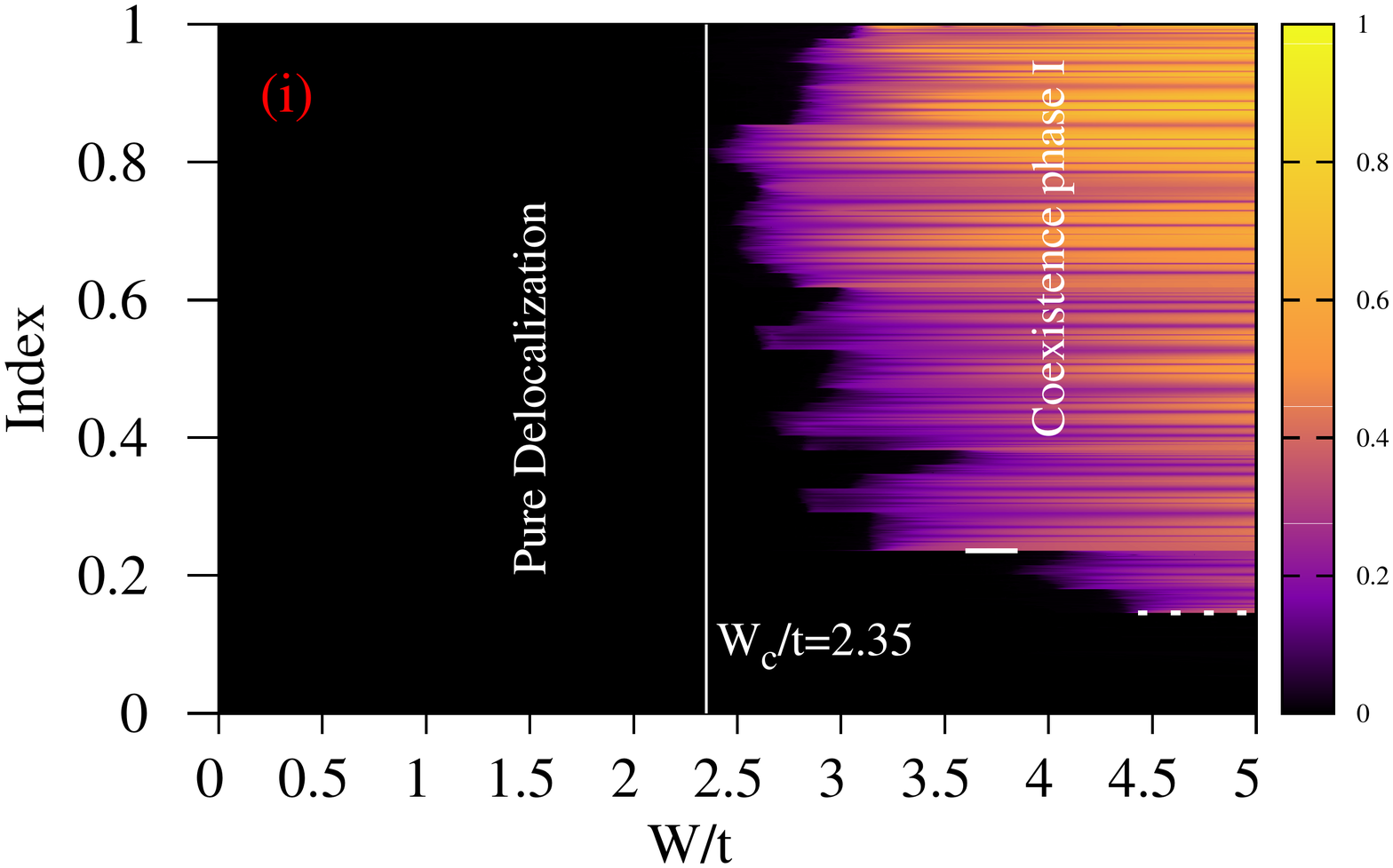}\hspace{-0.8cm}
\includegraphics[width=0.35\textwidth,height=0.30\textwidth]{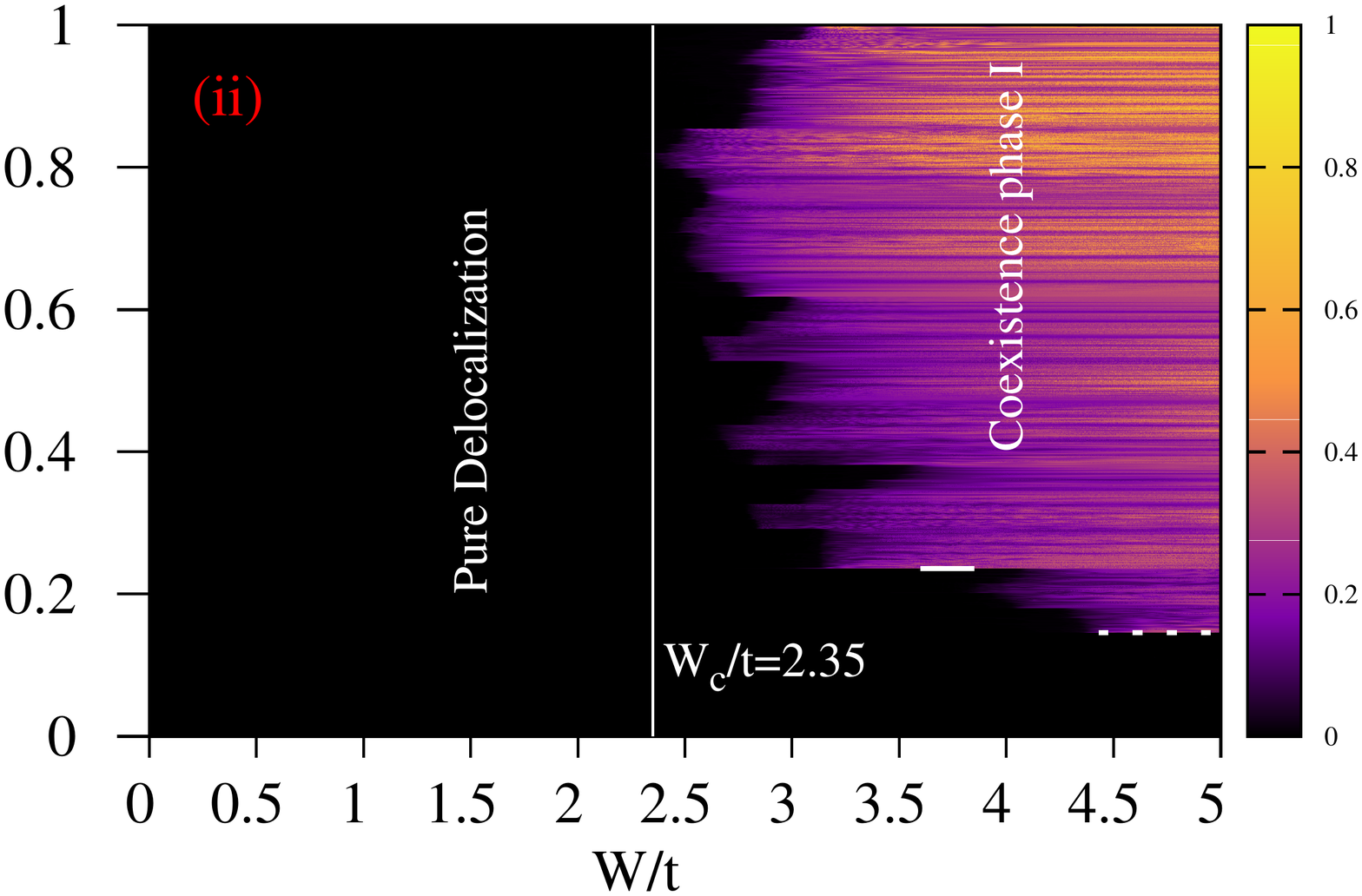}\hspace{-0.8cm}
\includegraphics[width=0.35\textwidth,height=0.30\textwidth]{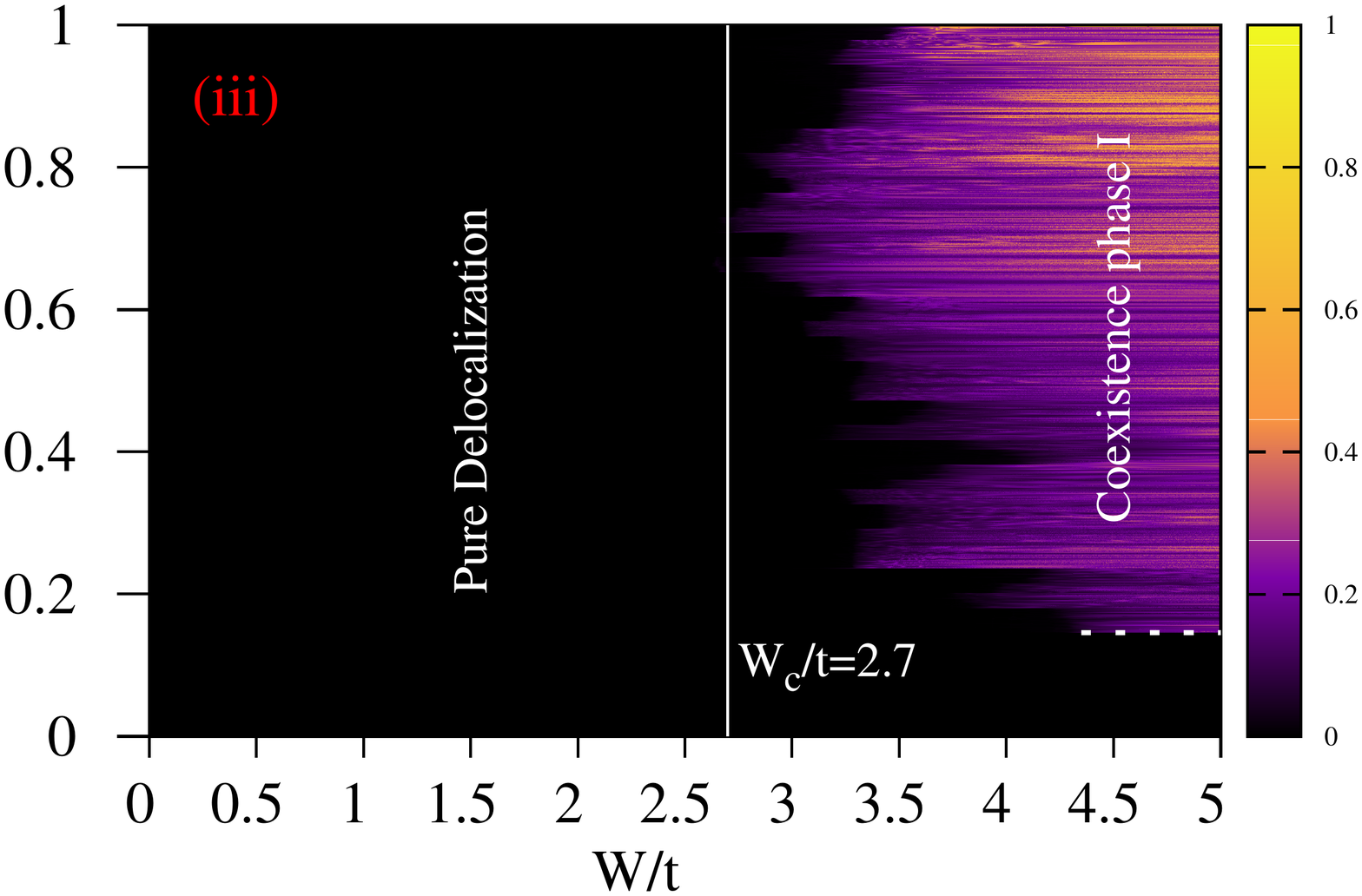}
\caption{Projection of IPR as function of $W/t$ and eigen-energies $(index)$ in the limit of the short-range $(a = 1.5)$ GAA model with RSO coupling . (i) $\alpha_y=0.8,$ $\alpha_z=0.0$, (ii) $\alpha_y=0.0,$ $\alpha_z=0.8$, and (iii) $\alpha_y=0.6,$ $\alpha_z=0.8$. Dotted lines show the presence of single mobilty edge regions.}
\label{Fig:IPR-SO-a1p5}
\end{figure*}
\begin{figure*}
\centering
\includegraphics[width=0.35\textwidth,height=0.30\textwidth]{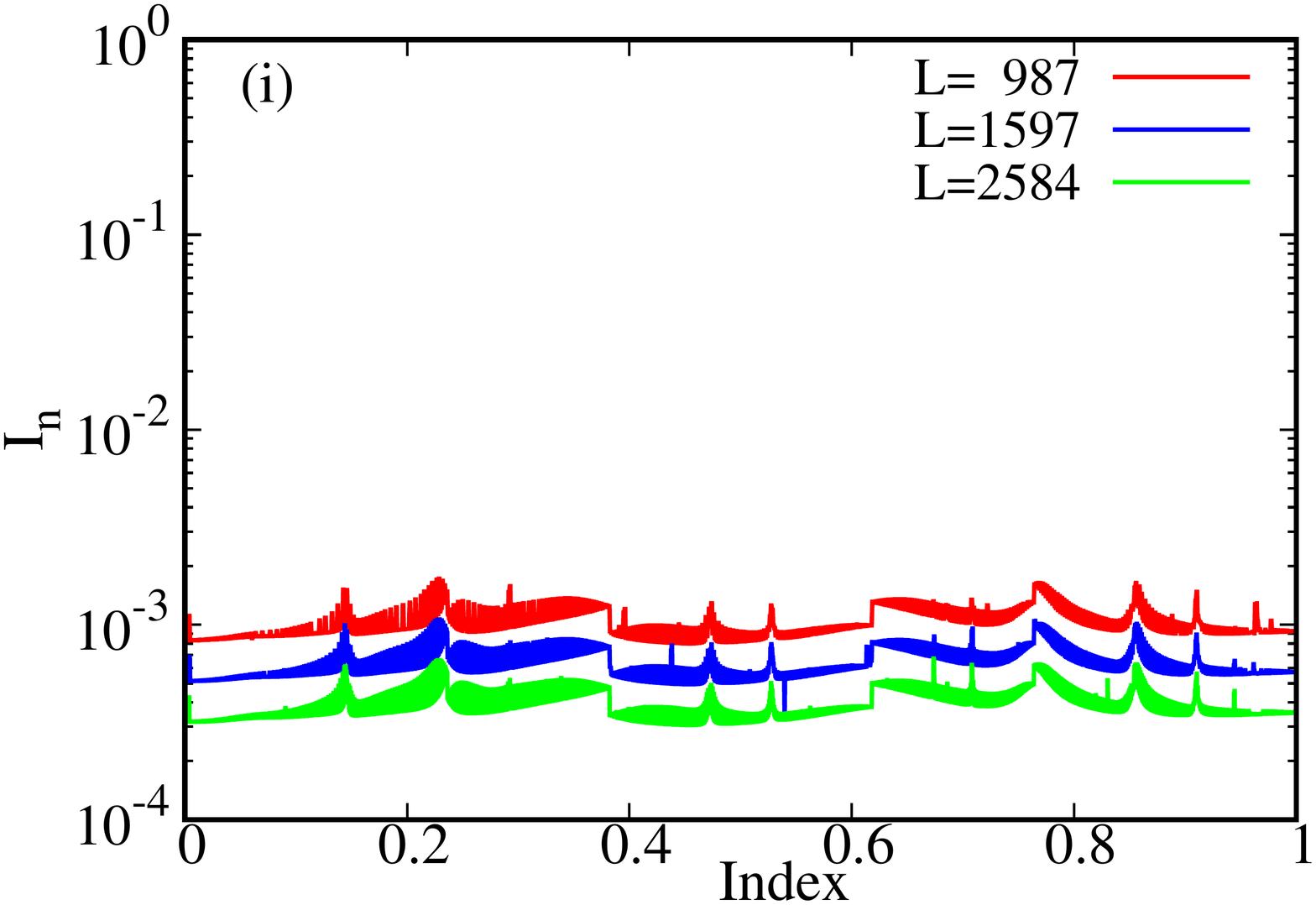}\hspace{-0.8cm}
\includegraphics[width=0.35\textwidth,height=0.30\textwidth]{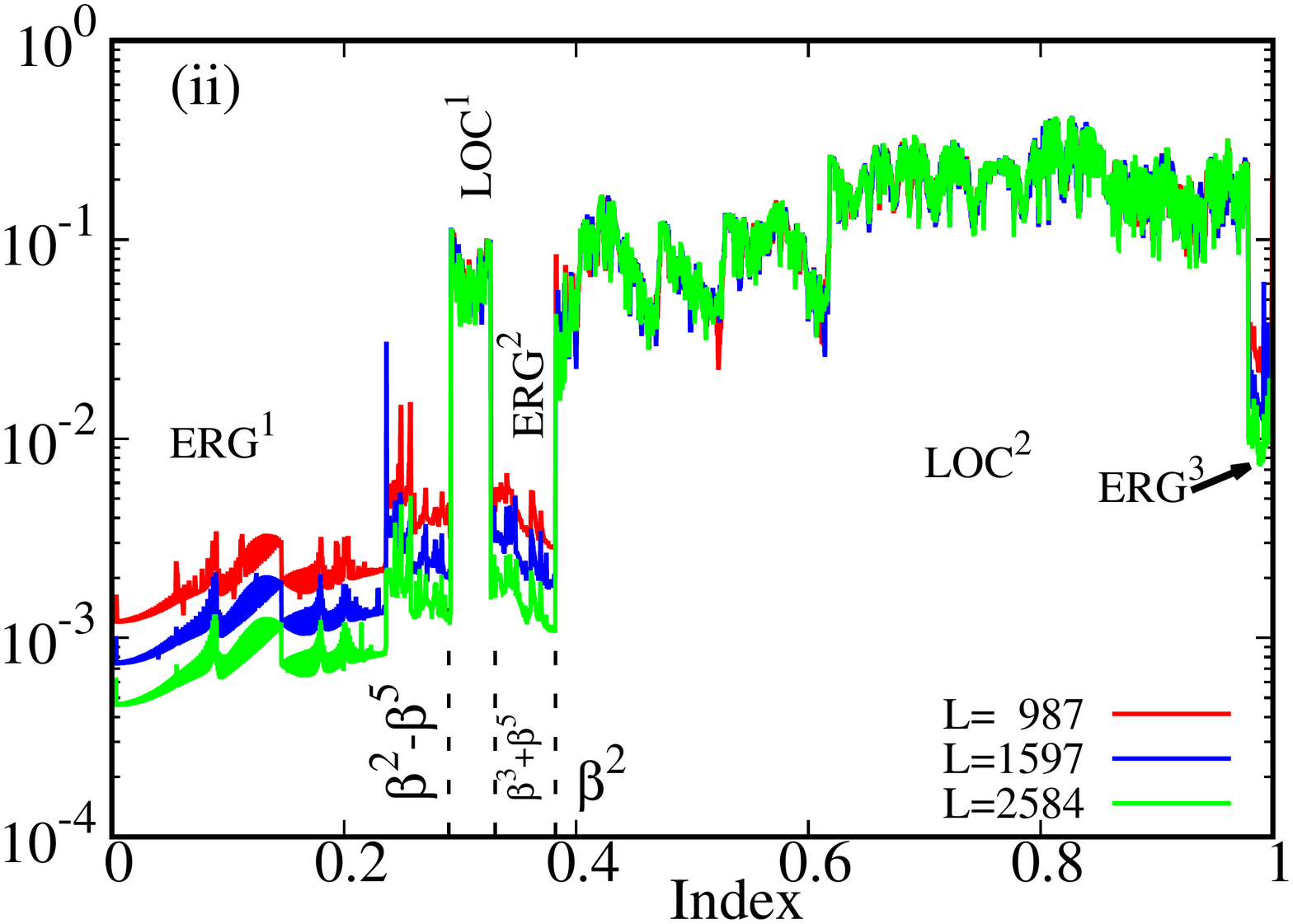}\hspace{-0.8cm}
\includegraphics[width=0.35\textwidth,height=0.30\textwidth]{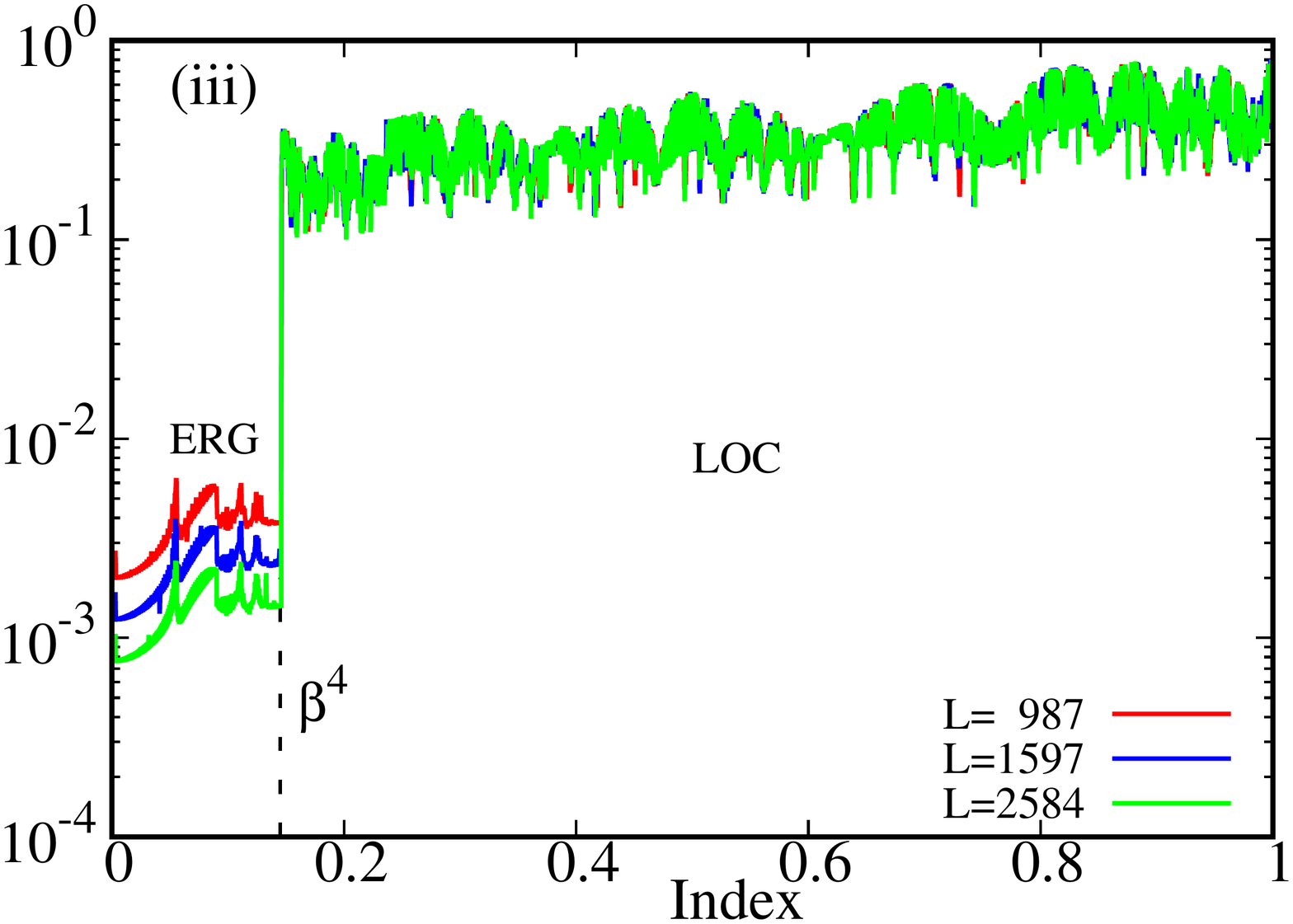}
\caption{IPR $(I_n)$ vs eigen-energies $(index)$ in the limit of the short-range ($a=1.5$) GAA model with RSO coupling at, (i) $W/t=1.2$, $\alpha_y=0.0,$ $\alpha_z=0.8$, (ii) $W/t=3.0$, $\alpha_y=0.0,$ $\alpha_z=0.8$, and (iii) $W/t=5.0$, $\alpha_y=0.0,$ $\alpha_z=0.8$.}
\label{Fig:IPR-SO-a1p5-LOC}
\end{figure*}

The energy spectrum changes dramatically in the limit of long-range hopping. Similar to the other limits, there exists a critical disorder strength below which all the eigenstates are delocalized. However, above the critical disorder strength a multifractal-edge separates delocalized 
and multifractal states. These results have been summarized in  Fig.~\ref{Fig:IPR-a0p5}(i). Here, the value of is $a$ is set to $0.5$. 
Apart from the appearance of the multifractal edge, the other crucial difference from the short-range hopping model is that all the states are multifractal type beyond the multifractal-edge. Furthermore, in contrast to the short range limit, there are no delocalized states in the energy spectrum above the multifractal edge. Interestingly, similar to the the short range hopping limit, in this case as well there exist disorder strength dependent sharp mobility edges in the spectrum. Furthermore, the location of the multifractal-edge follows exactly the same pattern as $a > 1$ limit that is the lowest $\beta^s L$ amount of states are delocalized depending on the strength of the disorder. To reveal the true nature of the states, we have plotted the IPR results with multiple system sizes in Figs.~\ref{Fig:IPR-a0p5}(ii) and (iii). In Fig.~\ref{Fig:IPR-a0p5}(ii), we have presented the IPR results for $W < W_c$, where $W_c/t \approx 0.4$. As expected, the IPR values scale inversely with the system size. The signature of multifractal states and multifractal edges can be seen in the IPR results presented in Fig.~\ref{Fig:IPR-a0p5}(iii).   
\begin{figure*}[ht]
\centering
\includegraphics[width=0.35\textwidth,height=0.30\textwidth]{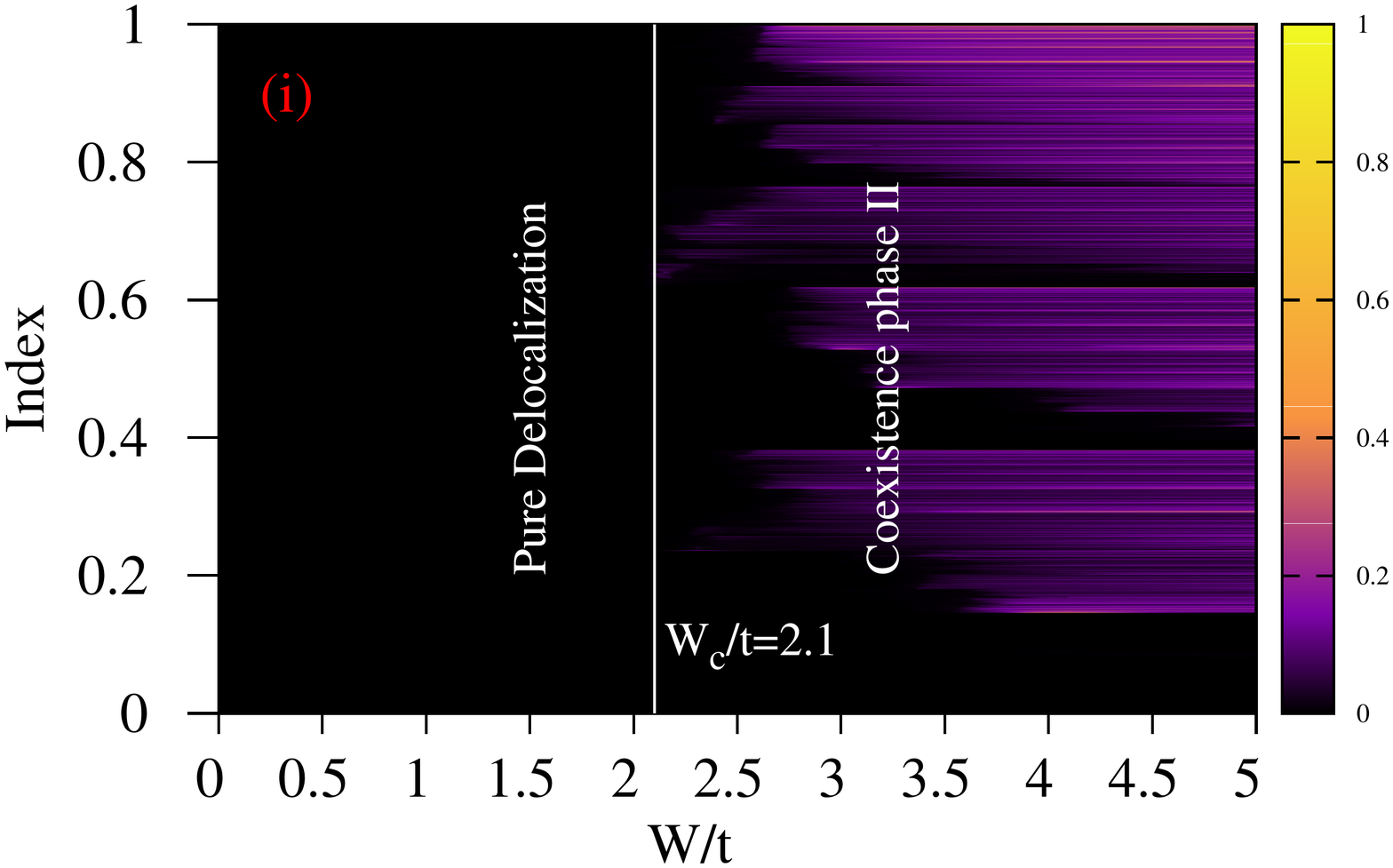}\hspace{-0.8cm}
\includegraphics[width=0.35\textwidth,height=0.30\textwidth]{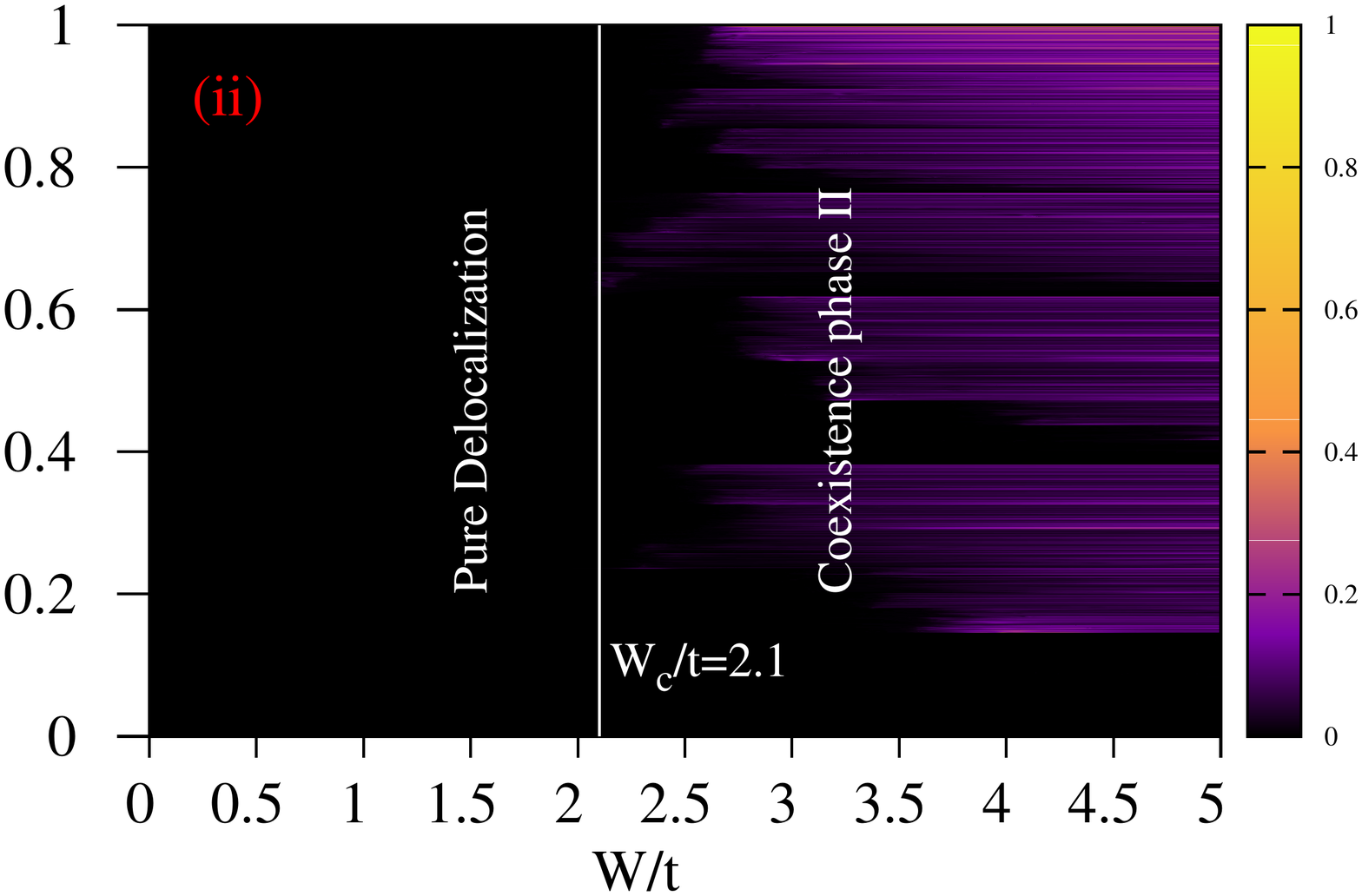}\hspace{-0.8cm}
\includegraphics[width=0.35\textwidth,height=0.30\textwidth]{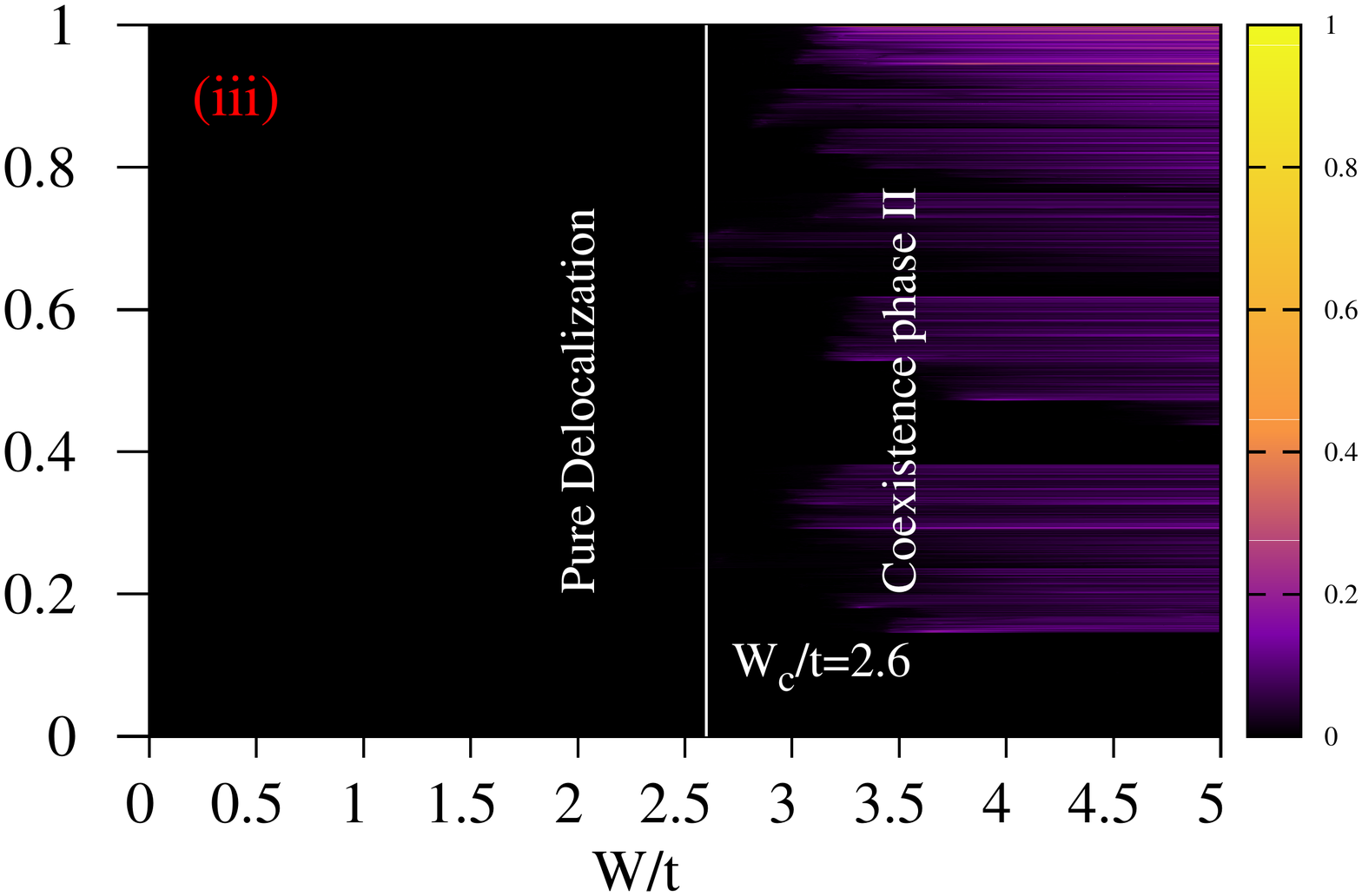}
\caption{Projection of IPR as function of $W/t$ and eigen-energies $(index)$ in the limit of the long-range $(a = 0.5)$ GAA model with RSO coupling . (i) $\alpha_y=0.8,$ $\alpha_z=0.0$, (ii) $\alpha_y=0.0,$ $\alpha_z=0.8$, and (iii) $\alpha_y=0.6,$ $\alpha_z=0.8$.}
\label{Fig:IPR-SO-a0p5}
\end{figure*}
\begin{figure*}[ht]
\centering
\includegraphics[width=0.35\textwidth,height=0.30\textwidth]{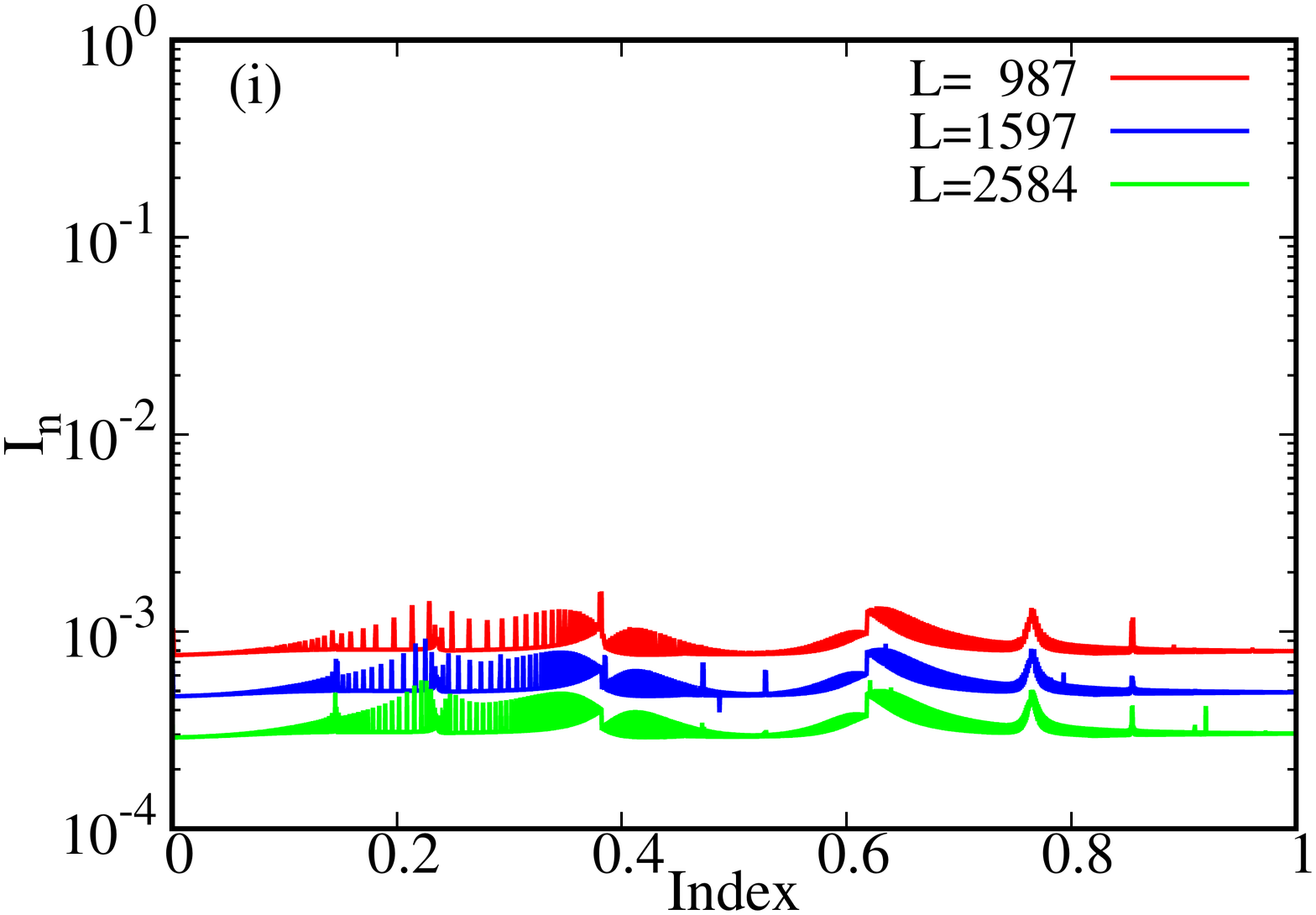}\hspace{-0.8cm}
\includegraphics[width=0.35\textwidth,height=0.30\textwidth]{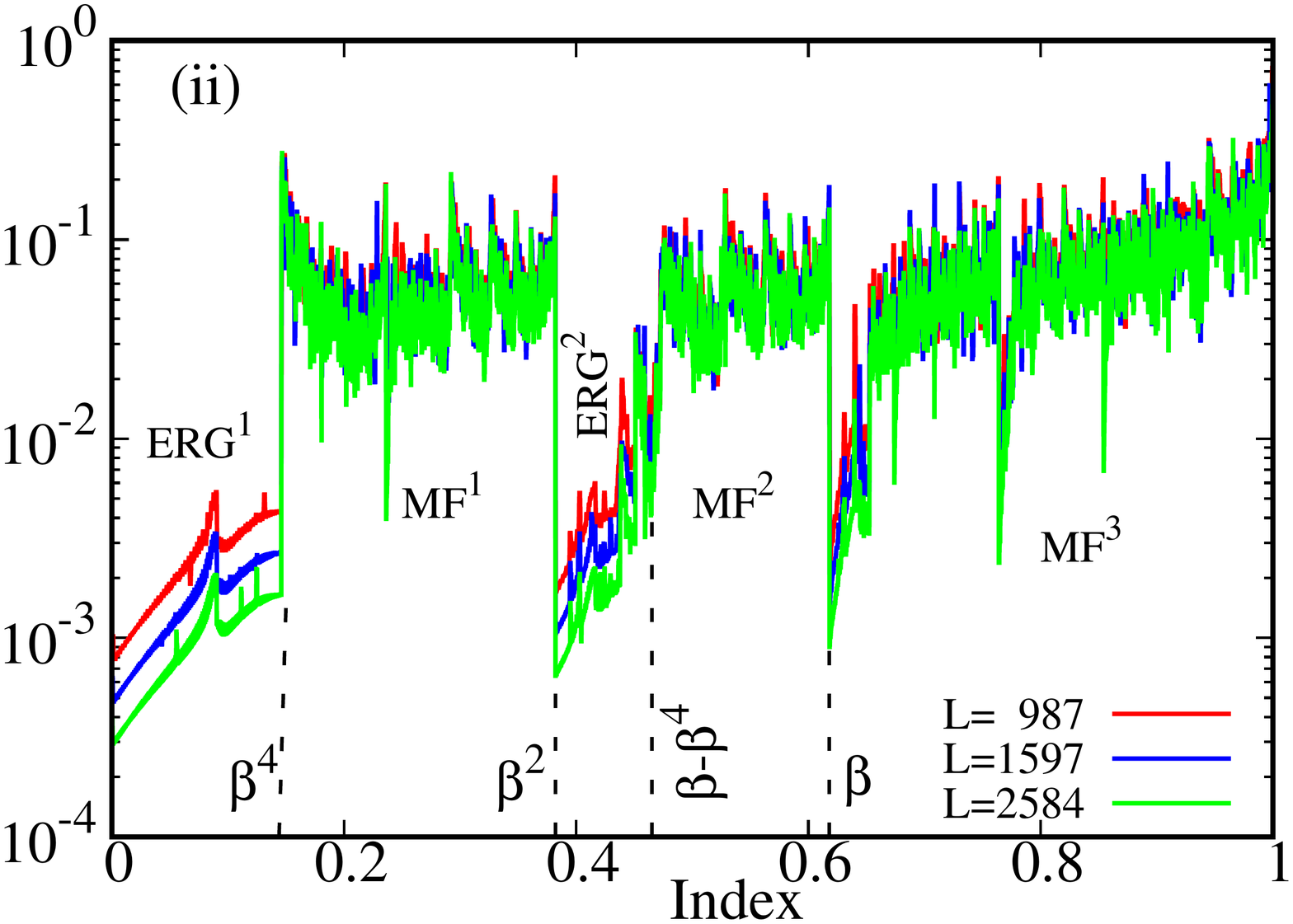}\hspace{-0.8cm}
\includegraphics[width=0.35\textwidth,height=0.30\textwidth]{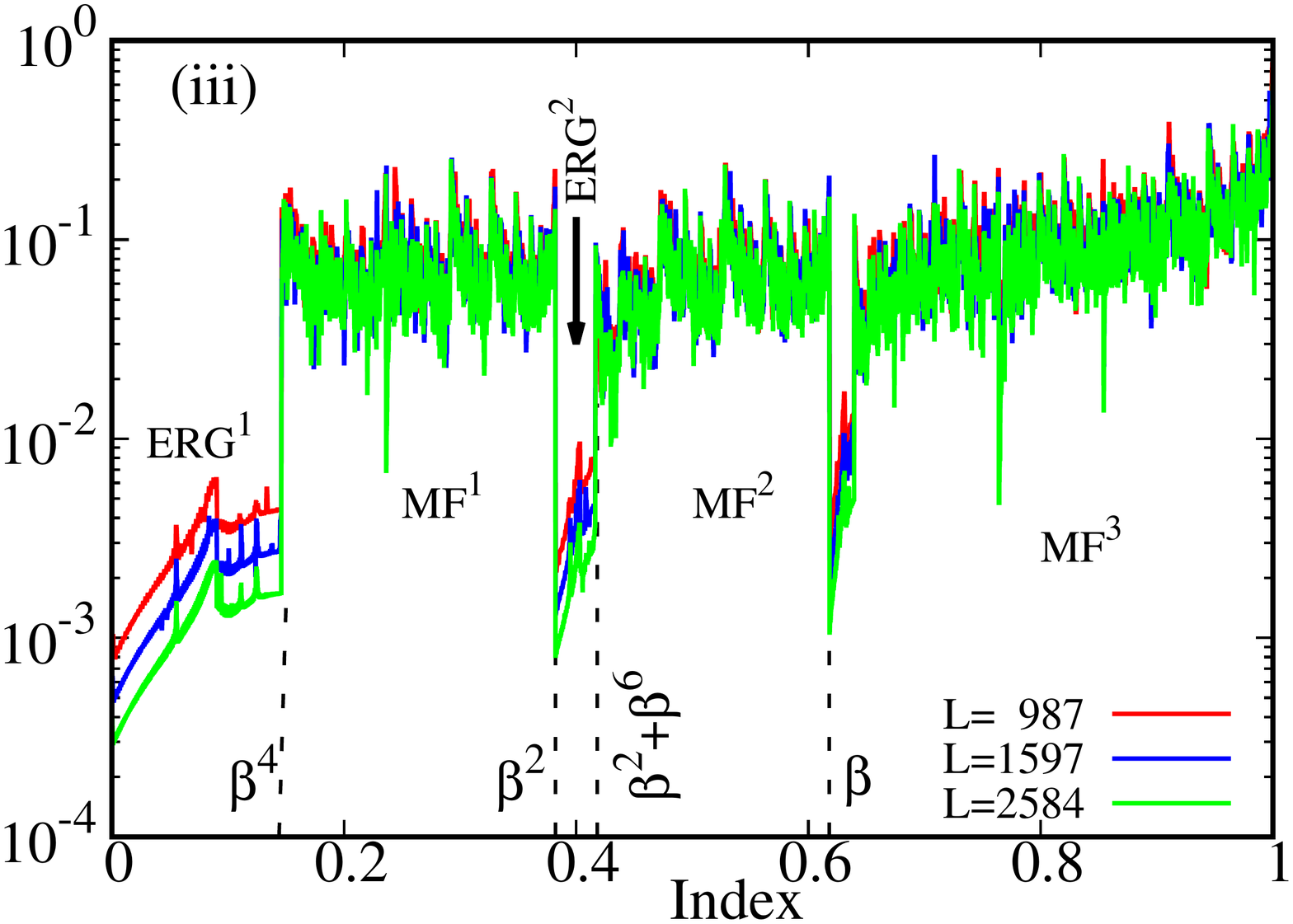}
\caption{IPR $(I_n)$ vs eigen-energies $(index)$ in the limit of the long-range ($a=0.5$) GAA model with RSO coupling at, (i) $W/t=0.5$, $\alpha_y=0.0,$ $\alpha_z=0.8$, (ii) $W/t=4.0$, $\alpha_y=0.0,$ $\alpha_z=0.8$, and (iii) $W/t=5.0$, $\alpha_y=0.0,$ $\alpha_z=0.8$.}
\label{Fig:IPR-SO-a0p5-MFS}
\end{figure*}
\subsection{GAA Hamiltonian with the RSO coupling}
We now discuss the effect of the RSO coupling on the energy spectrum of the GAA Hamiltonian. We are going to demonstrate that in the presence of RSO coupling, the energy spectrum of the GAA Hamiltonian dramatically changes, especially in the long range hopping limit. At first we discuss the effect of RSO coupling in the short-range hopping limit. In general, depending on the strength of the RSO coupling, the critical point shifts to a higher disorder strength irrespective of the range of the hopping. The effect of RSO coupling on the $a>>1$ limit has already been studied in Ref.~\cite{dsahu}. 

In Fig.~\ref{Fig:IPR-SO-a1p5}, we have analyzed the IPR result for the short-range $(a > 1)$ GAA model with RSO coupling. The value of $a$
is set to $1.5$, which is same as in Fig.~~\ref{Fig:IPR-a1p5}. It is quite evident that when the RSO coupling is introduced into the system the critical point shifts to a higher disorder strength. Without the RSO coupling, the critical disorder strength is $W_c/t \approx 1.1$. From Fig.~\ref{Fig:IPR-SO-a1p5}(i), we can observe that when only the spin-conserving part of the RSO Hamiltonian is turned on ($\alpha_y/t = 0.8, \alpha_z/t = 0$) the critical disorder strength turns out to be $W_c/t \approx 2.4$. Interestingly, the spin-flip hopping has an identical effect on the critical disorder strength. In Fig.~\ref{Fig:IPR-SO-a1p5}(ii) we have presented the IPR results as a function of $W/t$ for the case when only the spin-flip hopping part of the RSO Hamiltonian is turned on. In this case $\alpha_y/t = 0$ while $\alpha_z/t = 0.8$. It is clear 
that the critical disorder strength turns out to be $W_c/t \approx 2.4$ in this case as well. Moreover, the individual effect of the two types of hopping of the RSO Hamiltonian on the energy spectrum is also identical. 
A comparison between the IPR results of Fig.~\ref{Fig:IPR-a1p5}(i) and Fig.~\ref{Fig:IPR-SO-a1p5}(i) reveal that in the presence of the RSO coupling the disorder windows at $\beta$ and $\beta^2$ of the GAA Hamiltonian have now disappeared. The mobility edge, where all the states are localized above it, now first appears at $\beta^3$, then at $\beta^4$ with an increase in the disorder strength. Outsidethese windows, in the presence of either of the RSO couplings, multiple mobility edges 
can be found where small bands of delocalized states separate bands of localized states. The effect of a weaker RSO coupling strength is qualitatively similar, and in Appendix~\ref{Sec:RSO-IPR-Small} we have presented some results to demonstrate this.

To demonstrate the existence of multiple mobility edges beyond the critical point, in Fig.~\ref{Fig:IPR-SO-a1p5-LOC} we have plotted the 
IPR results as a function of the eigeneneries below the critical point, and for two different disorder strengths above the critical value. From 
Fig.~\ref{Fig:IPR-SO-a1p5-LOC}(i) ($W/t = 1.2$), it is clear that the IPR values across the entire energy spectrum scale inversely with the system size indicating that all the states are delocalized. The most interesting features can be observed at $W/t=3$ (Fig.~\ref{Fig:IPR-SO-a1p5-LOC}(ii)). It is evident from this figure that a mobility edge first appears approximately at $n/L=\beta^2-\beta^5$, below which all the states are delocalized, and above which a narrow band of localized states appear. These localized states survives till $n/L=\beta^3 + \beta^5$ where another mobility edge appears beyond which another narrow band of delocalized states can be observed till $n/L=\beta^2$. After this point, the states are localized till the appearance of a tiny set of delocalized states close to highest eigenenergies of the spectrum. When the disorder strength is set at $W/t=5.0$ in Fig.~\ref{Fig:IPR-SO-a1p5-LOC}(iii), only a single mobility edge at $n/L=\beta^4$, originally present in the pure GAA Hamiltonian, can be seen in the energy spectrum.

As quite expected, when both the RSO coupling is present, the critical disorder strength is pushed towards a higher disorder strength compared to the case when only one of the RSO coupling is considered. In Fig.~\ref{Fig:IPR-SO-a1p5}(iii), the strength of the spin-flip hopping amplitude $\alpha_z/t$ has been fixed at $0.8$ as in Fig.~\ref{Fig:IPR-SO-a1p5}(ii) while $\alpha_y/t$ is set to value of $0.6$. It is quite evident that the critical disorder strength in this case is $W_c/t \approx 2.7$. We would like to point out that the critical point and the spectrum remain unchanged if the strengths of $\alpha_y/t$ and $\alpha_z/t$ are
interchanged. The presence of both the RSO couplings do not introduce any significant change in the qualitative behaviour of the energy spectrum, except that the window of disorder strengths at which only a single mobility edge exists is now appear first at $n/L=\beta^4$.  

The effect of the RSO coupling on the energy spectrum is more dramatic in the case of the long-range hopping limit $(a \leq 1)$ of the GAA Hamiltonian. In this limit as well, for a given value of $a$, the critical disorder strength increases in the presence of the RSO coupling compared to the pure GAA Hamiltonian. Similar to the short-range hopping limit case, when either of the two hopping processes of the RSO Hamiltonian is considered, the critical disorder strength turns out to be same for the same strength of $\alpha_y/t$ or $\alpha_z/t$. This can be observed from the results presented in Figs.~\ref{Fig:IPR-SO-a0p5}(i) and \ref{Fig:IPR-SO-a0p5}(ii). However, the effect of the RSO couplings on the spectrum is remarkably different in the case of long-range hopping limit. It can be observed from Fig.~\ref{Fig:IPR-SO-a0p5}(i) (Fig.~\ref{Fig:IPR-SO-a0p5}(ii)) that when $\alpha_y/t = 0.8$ and $\alpha_z/t = 0.0$ ($\alpha_y/t = 0.0$ and $\alpha_z/t = 0.8$) the spectrum no longer has a single multifractal edge at any disorder strength, rather multiple multifractal edges emerge beyond the critical point, which separates successive bands of delocalized and multifractal states. The regular pattern of the lowest $\beta^s L$ amount states being delocalized in the case of pure GAA Hamiltonian is no longer true. However, the location of the multifractal edges (separating the delocalized and multifractal states or vice versa) are in general approximately given by $n/L=\beta^s \pm \beta^m$, where $s,m=0,1,2,\cdots$.This is quite similar to the short-range GAA Hamiltonian in the presence of the RSO coupling discussed in the previous paragraph. 

To demonstrate the existence of multiple multifractal edges and their locations, in Fig.~\ref{Fig:IPR-SO-a0p5-MFS} we have plotted the IPR values as a function of the eigeneneries below and above the critical point. We have set  $\alpha_y/t =0$ and $\alpha_z/t =0.8$. From Fig.~\ref{Fig:IPR-SO-a0p5-MFS}(i), it is evident that below the critical disorder strength all the eigenstates are delocalized. From Figs.~\ref{Fig:IPR-SO-a0p5-MFS}(ii) and (iii), however, we can see that above the critical disorder strength the spectrum gets divided into alternative regions of multifractal and delocalized states. For $W/t=4.0$, in the GAA Hamiltonian without the RSO coupling, a single multifractal edge appears at $n/L= \beta^4$. From Fig.~\ref{Fig:IPR-SO-a0p5-MFS}(ii), however, it is clear that in the presence of the RSO coupling, bands of delocalized states emerges between $n/L=\beta^2$ and $n/L=\beta-\beta^4$, and again at $n/L=\beta$. From Fig.~\ref{Fig:IPR-SO-a0p5-MFS}(iii) it is also quite evident that with an increase in the disorder strength, the width of these delocalized bands shrink. 

From the IPR results, the existence of multiple mobility or multifractal edges in the GAA Hamiltonian in the presence of RSO coupling is quite clear. Furthermore, we can identify the localized and multifractal states from these results only qualitatively. A more quantitative approach is needed to analyze the states to distinguish the localized and the multifractal states unambiguously. In the next section, we present the multifractal analysis of the quasiparticle eigenstates, which helps to analyze the states quantitatively.

\begin{figure*}
\centering
\includegraphics[width=0.26\textwidth,height=0.28\textwidth]{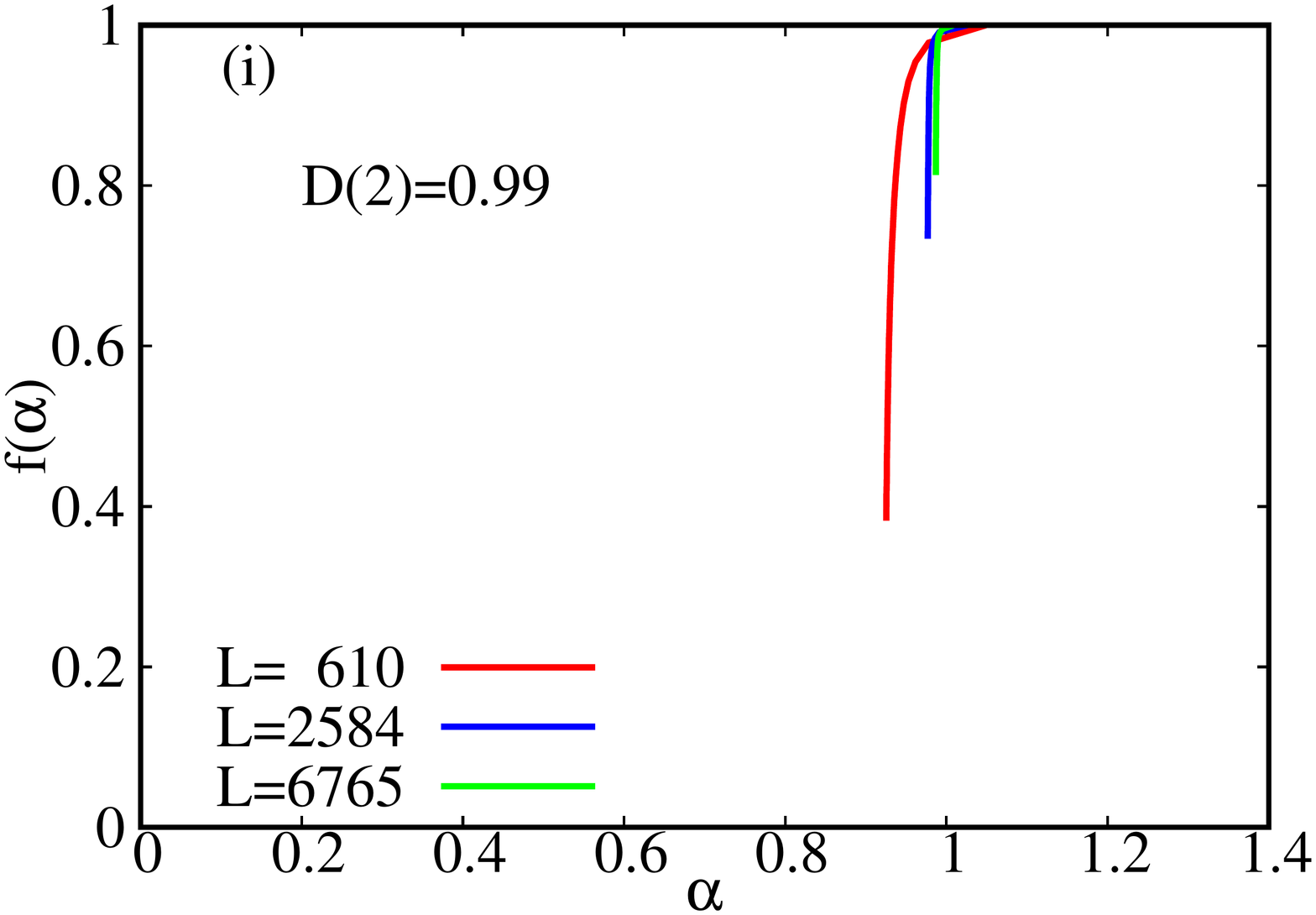}\hspace{-0.8cm}
\includegraphics[width=0.26\textwidth,height=0.28\textwidth]{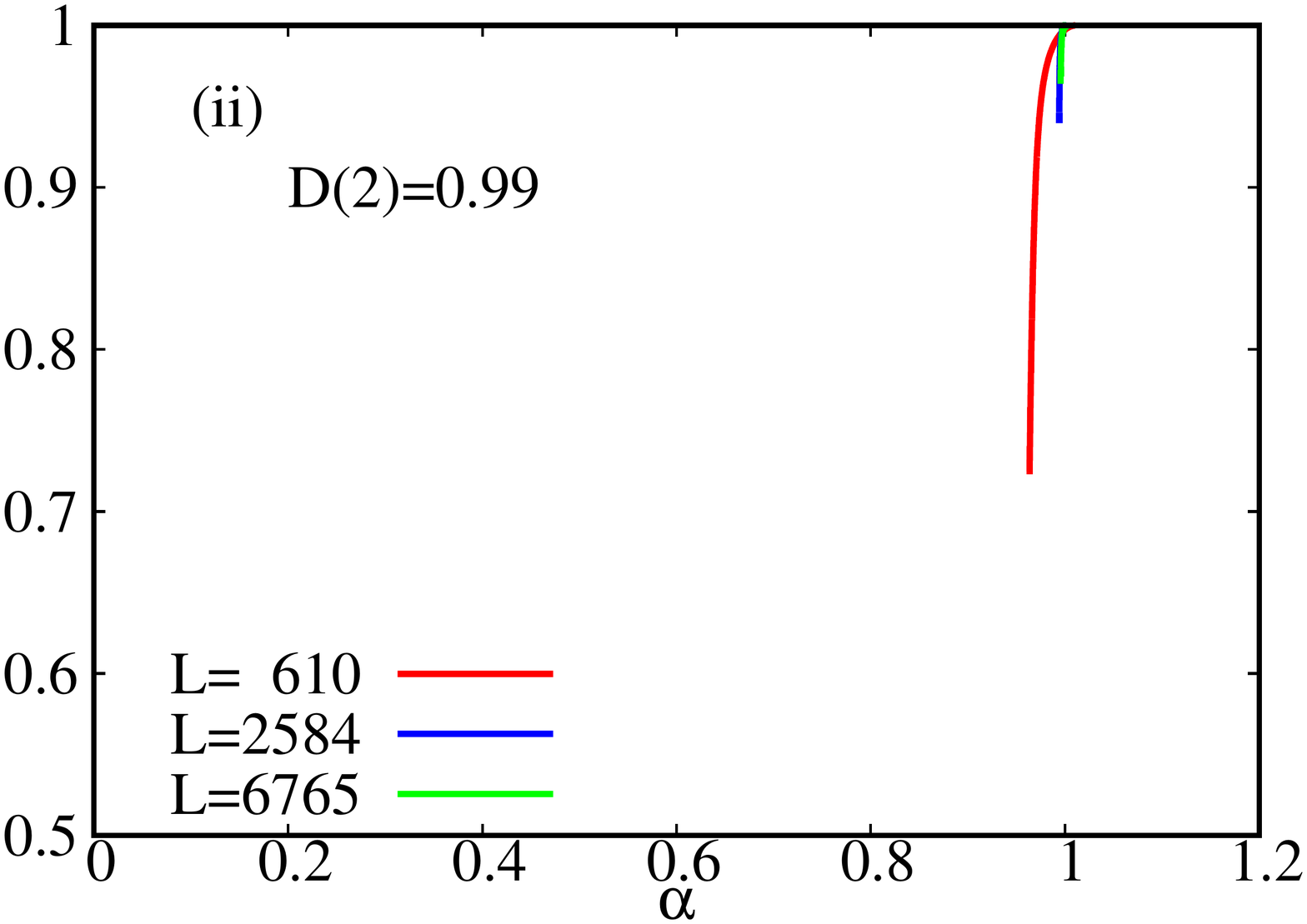}\hspace{-0.8cm}
\includegraphics[width=0.26\textwidth,height=0.28\textwidth]{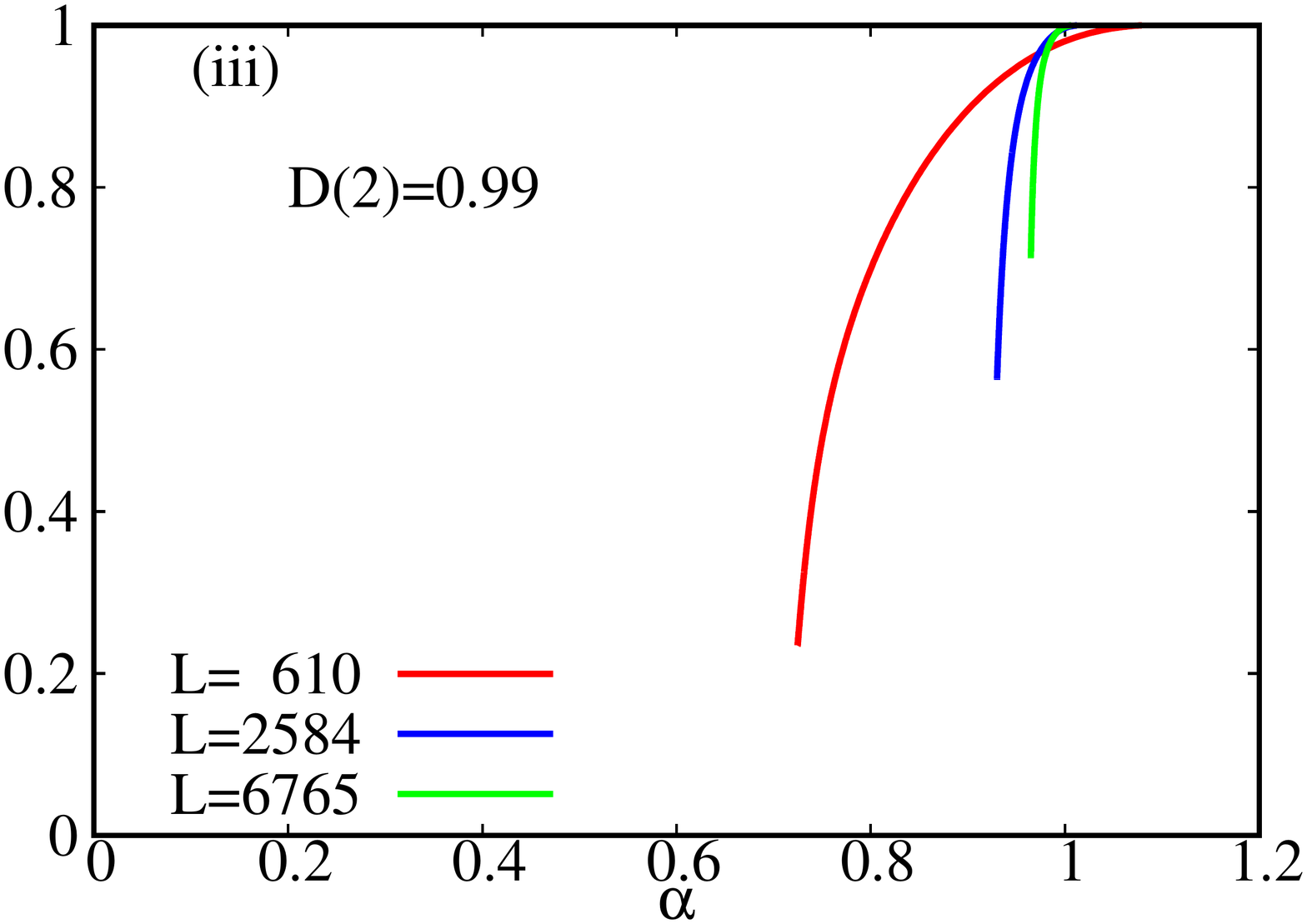}\hspace{-0.8cm}
\includegraphics[width=0.26\textwidth,height=0.28\textwidth]{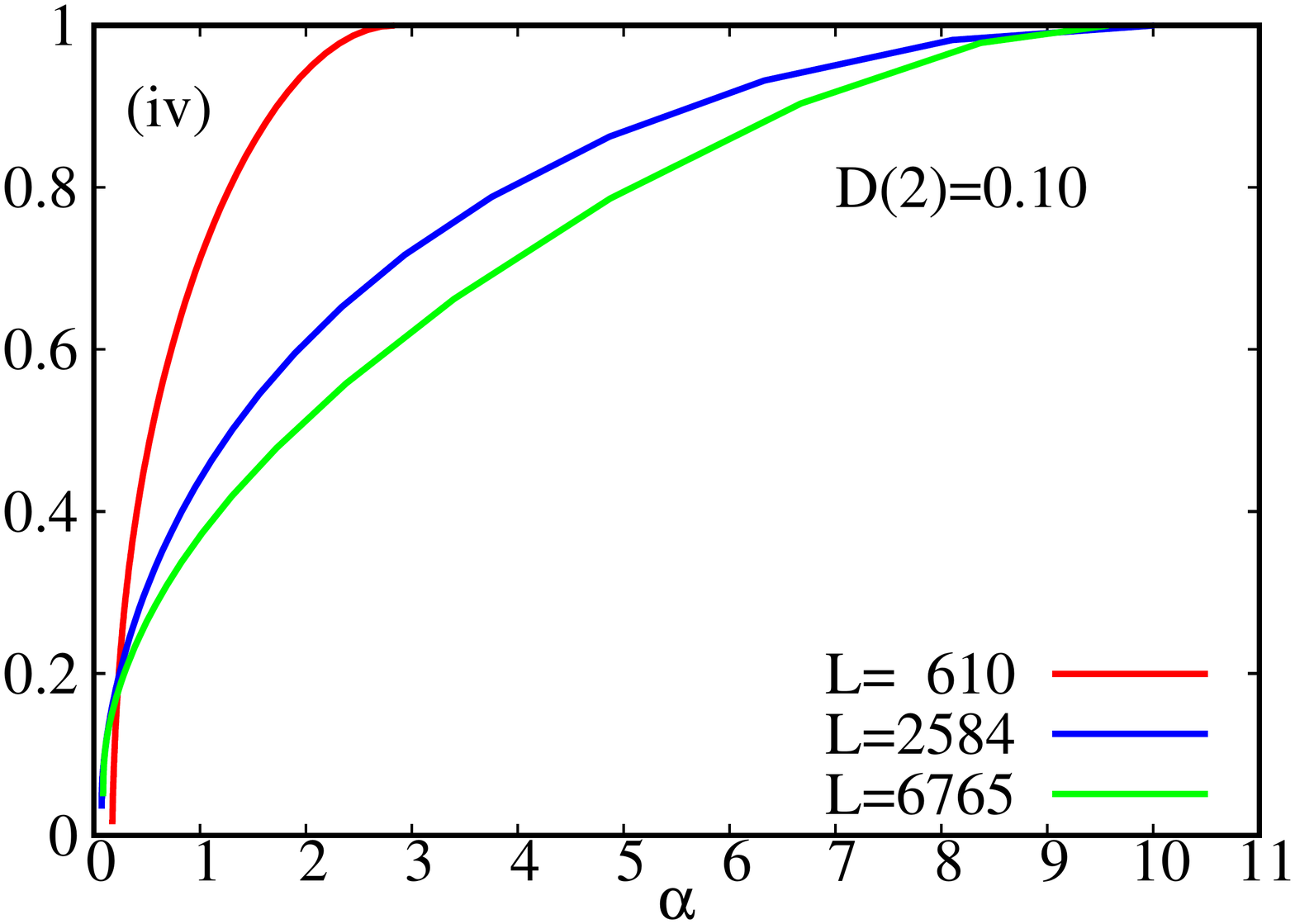}
\caption{Multifractal analysis of the short-range ($a=1.5$) GAA model with RSO coupling coefficients as $\alpha_y=0$,$\alpha_z = 0.8$. (i) at $W/t=1.2$ (Pure delocalization region). (ii) at $W/t=3.0$ in the $ERG^1$ 
region of Fig.\ref{Fig:IPR-SO-a1p5-LOC}(ii). (iii) at $W/t=3.0$ in the $ERG^2$ region of Fig.\ref{Fig:IPR-SO-a1p5-LOC}(ii). (iv) at $W/t=3.0$ in the $LOC^1$ region of Fig.\ref{Fig:IPR-SO-a1p5-LOC}(ii).}		
\label{Fig:MFS-1p5}
\end{figure*}

\section{Multifractal Analysis of Eigenstates}\label{Sec:multifractal-spectrum}
The IPR results show the fragmentation of the energy spectrum into successive windows of delocalized, localized, or multifractal states in the 
presence of RSO coupling in the GAA model. In this section we carry out the multifractal analysis, especially for $W>W_c$, to determine the true nature of the quasiparticle eigenstates in the short and long range limits. 
In the multifractal analysis \cite{Halsey,Martin Janssen}, one generally starts by identifying that 
$P_{n}(q) = \sum_{i=1}^{N} \left|\psi_n(i) \right|^{2 q} \propto N^{-\tau(q)}$, where $\psi_n$ is the normalized wave function corresponding 
to $n$-th eigenvalue and $i=1,2,\cdots,N$. The exponent $\tau(q)$ is alternatively expressed in terms of $D(q)$ as
\begin{equation}
\tau(q) = D(q)/(q-1), 
\end{equation}
where $D(q)$ is called the fractal dimension. 

Typically, to understand the nature of the eigenstates the most commonly used quantity is $D(2)$, as its numerical value indicates the degree 
of multifractality. It turns out that $\tau(q) = q-1$ in case of delocalized states \cite{DeLuca}, which effectively means that for that for delocalized states $D(2) = 1$. In the other extreme case, when the eigenstates are completely localized $D(2) = 0.$ For the multifractal states $D(2)$ lies in between these two limits, that is $0 < D(2) < 1$. Hence, in principle if one can compute $\tau(q)$ then in principle $D(q)$ can be estimated. In practice, however, instead of directly computing $\tau(q)$ an equivalent parameter $f(\alpha)$ is computed, which characterizes the multifractal property of the eigenstates much more
conveniently. These two parameters $f(\alpha)$ and $\tau(q)$ are connected by the Legendre transformation as follows,
\begin{equation}
f(\alpha(q)) = q \alpha(q) -\tau(q), 
\end{equation}
where $\alpha(q) = d \tau(q)/dq$. 

Generally, $f(\alpha)$ is a smooth positive valued convex function of $\alpha$. It has a global maximum but no local minima or maxima. It is interesting to note that, $f_{max} = f(\alpha(q=0)) = d$, where $d$ is the Euclidean dimension of the system \cite{Martin Janssen}. From the $f(\alpha)$ vs $\alpha$ spectrum, it is easy to identify the nature of the eigenstates. For delocalized states, when $\alpha \neq 1 $, 
$f(\alpha)= -\infty$ and $f(\alpha=1)= 1$ in the limit of large system sizes. For multifractal states $f(\alpha(q))$ is bounded and non-zero only 
between $0< \alpha_{min} < \alpha(q) < \alpha_{max}$, while for the localized states $0=\alpha_{min} < \alpha(q) < \alpha_{max}$, where $\alpha_{max} = \alpha(q=0)$.
\begin{figure*}
\centering
\includegraphics[width=0.26\textwidth,height=0.28\textwidth]{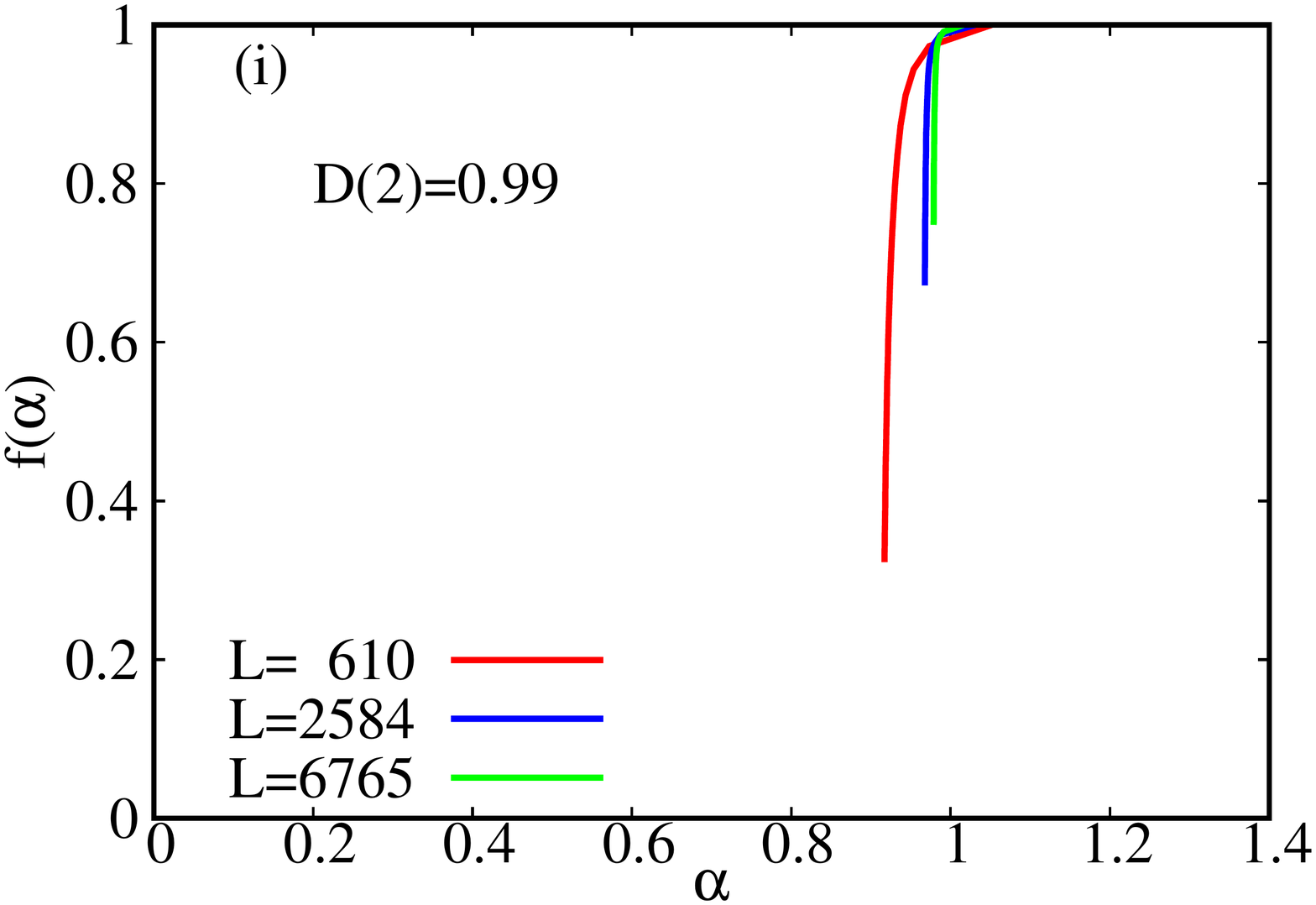}\hspace{-0.8cm}
\includegraphics[width=0.26\textwidth,height=0.28\textwidth]{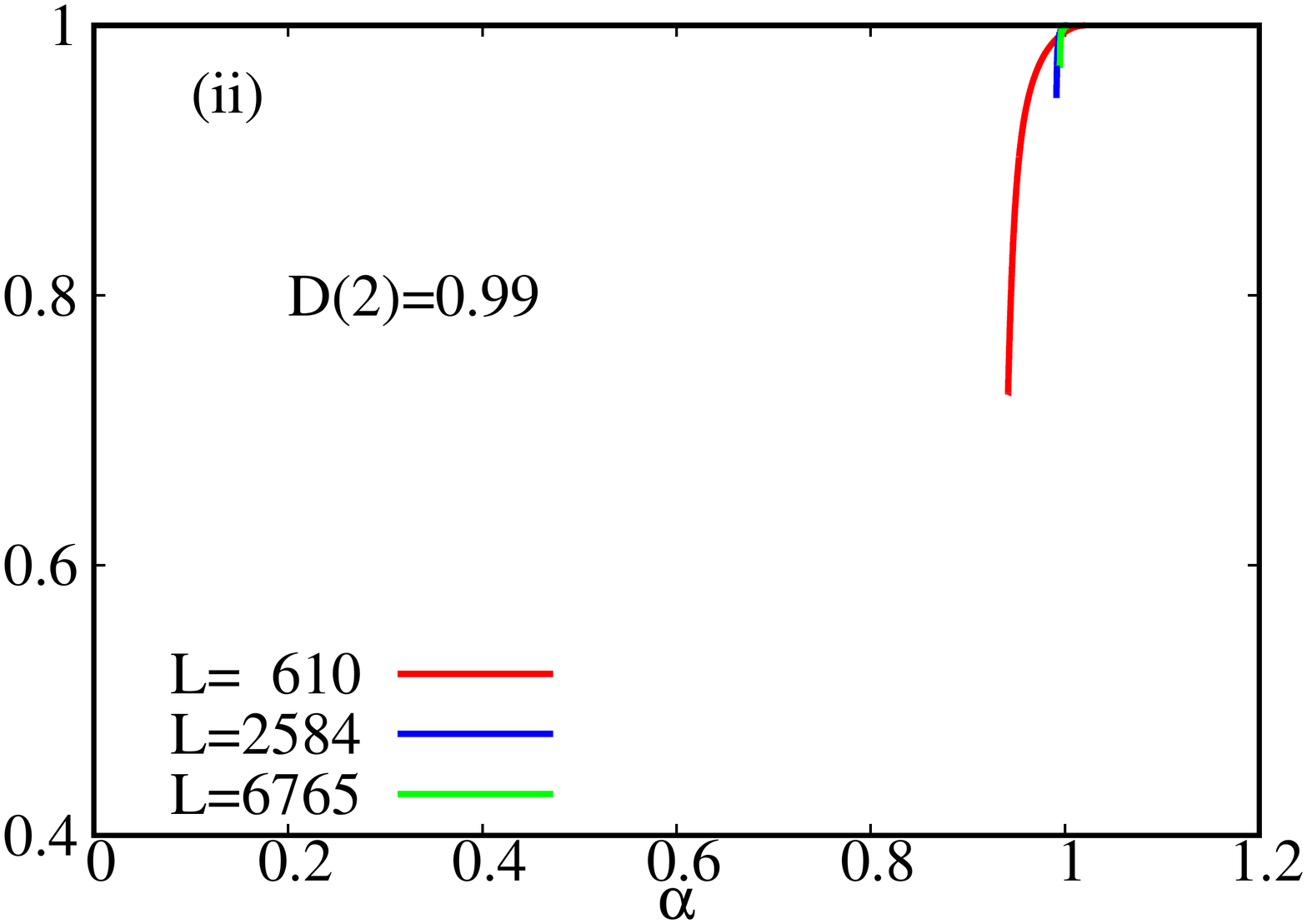}\hspace{-0.8cm}
\includegraphics[width=0.26\textwidth,height=0.28\textwidth]{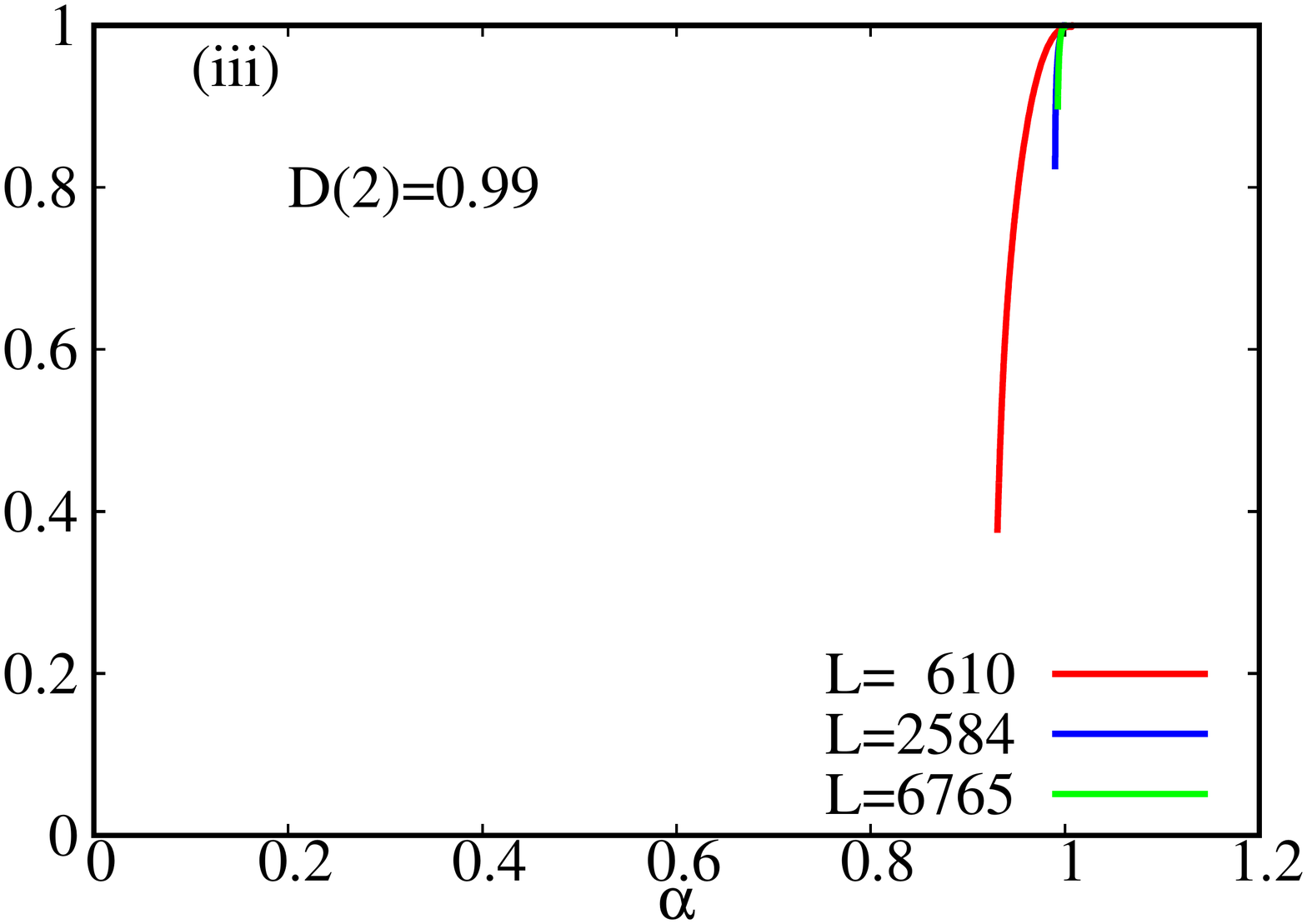}\hspace{-0.8cm}
\includegraphics[width=0.26\textwidth,height=0.28\textwidth]{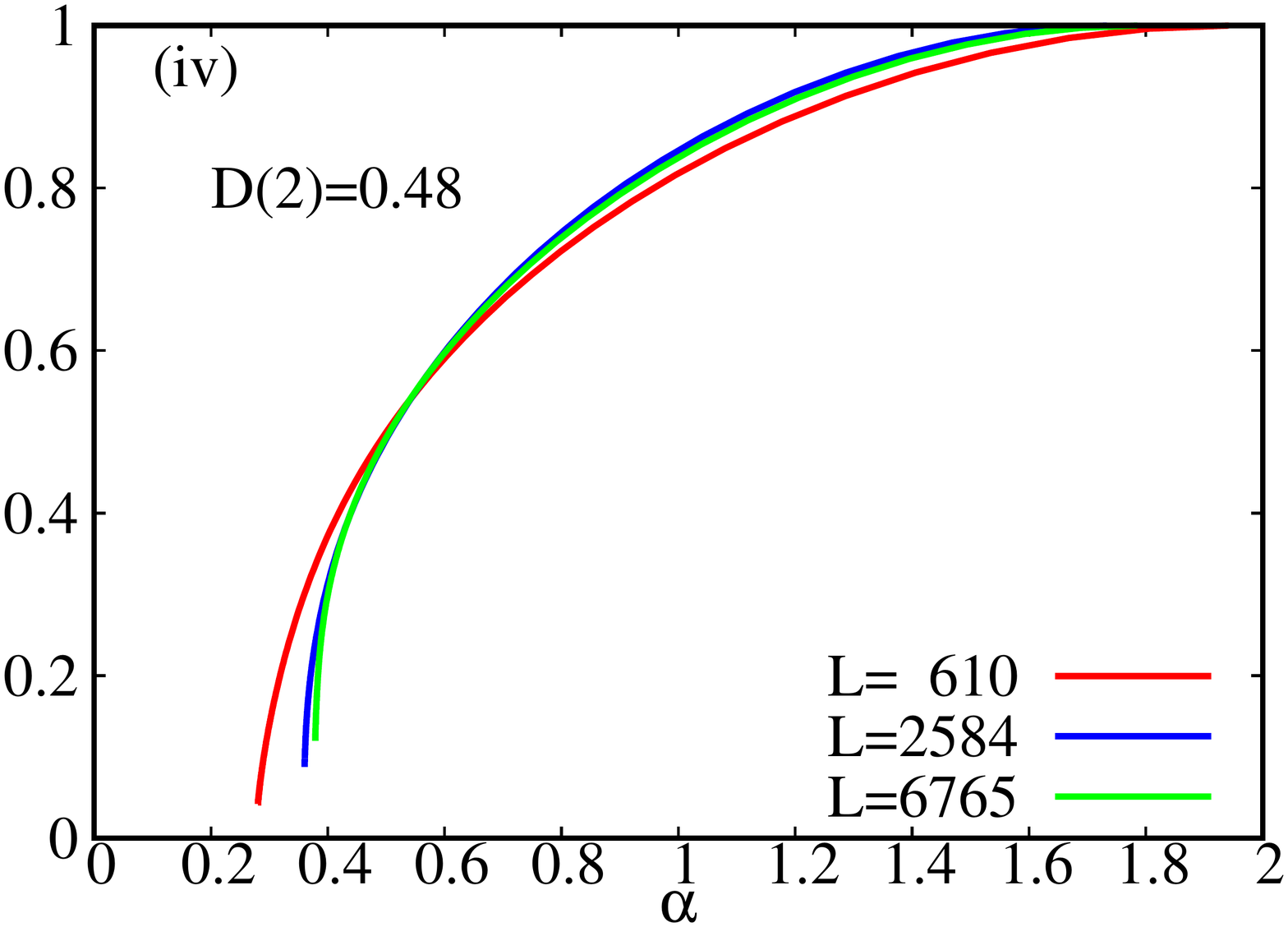}
\caption{ Multifractal analysis of the long-range ($a=0.5$) GAA model with RSO coupling coefficients as $\alpha_y=0$,$\alpha_z = 0.8$. (i) at $W/t=1.0$ (Pure delocalization region). (ii) at $W/t=4.0$ in the $ERG^1$ region of Fig.\ref{Fig:IPR-SO-a0p5-MFS}(ii). (iii) at $W/t=4.0$ in the $ERG^2$ region of Fig.\ref{Fig:IPR-SO-a0p5-MFS}(ii). (iv) at $W/t=4.0$ in the $MF^1$ region of Fig.\ref{Fig:IPR-SO-a0p5-MFS}(ii).}	
\label{Fig:MFS-0p5}
\end{figure*}
\subsection{Calculation of Multifractal Spectrum}
From the previous discussion we observe that to analyze the nature of the eigenstates it is sufficient to compute the multifractal spectrum $f(\alpha(q))$ for $0 \leq \alpha(q) \leq \alpha(q=0)$. In this article, we are going to follow the Ref.\cite{Martin Janssen} and Ref.\cite{Cuevas} to numerically compute the multifractal function $f(\alpha(q))$. Before discussing our results, we briefly review the key steps to determine $f(\alpha(q))$. Initially the lattice is divided into small boxes of linear size $l < L$, where $L$ is the length of the lattice, and then we compute the following normalized probability,
\begin{equation}
\mathcal{P}_{k}(l,q)=\frac{\mathcal{P}_{k}^{q}(l)}{\sum_{j=1}^{N_b}\mathcal{P}_{j}^{q}(l)},  
\end{equation}
where $1\leq k \leq N_b=L/l$ represents the $k$-th box and $\mathcal{P}_{k}^{q}(l)=\sum_{i \in l_k} \left|\psi_n(i)\right|^{2}, l_k = l ~ \forall ~ k ,$ is the probability 
of the $n$-th eigenstate. Once $\mathcal{P}_{k}^{q}(l)$ is computed, $\alpha(q,L)$ and $f(\alpha(q,L))$ can be obtained from the following relations;
\begin{eqnarray}
\alpha(q,L) &=& \underset{\delta \rightarrow 0}{\mathrm{lim}} 
\frac{\sum_{k=1}^{N_b}\mathcal{P}_{k}(l,q) \textrm{ln}(\mathcal{P}_{k}(l,1))}{\textrm{ln} \delta}  \\
f(\alpha(q,L)) &=& \underset{\delta \rightarrow 0}{\mathrm{lim}} 
\frac{\sum_{k=1}^{N_b}\mathcal{P}_{k}(l,q) \textrm{ln}(\mathcal{P}_{k}(l,q))}
{\textrm{ln} \delta} ,
\end{eqnarray}
where $\delta = l/L$.

Here, we mainly focus on the multifractal analysis of the narrow band of states that appear beyond the critical point in the presence of the 
RSO coupling. We have presented our results for $\alpha_y=0 $, and $\alpha_z\neq0$. In Fig.~\ref{Fig:MFS-1p5}, we have presented the multifractal spectrum for the short-range $(a > 1)$ GAA Hamiltonian in the presence of RSO coupling. At first, in Fig.~\ref{Fig:MFS-1p5}(i), the multifractal spectrum is presented for the disorder strength $W/t=1.2$, which is below the critical value of $W_c/t \approx 2.35$. From the results it is clear that as the system size increases to a large value $f(\alpha=1)$  converges to the value $1$, which is the signature of purely delocalized states. This result corroborates our observations made from the IPR results of Fig.~\ref{Fig:IPR-SO-a1p5}(i) and Fig.~\ref{Fig:IPR-SO-a1p5-LOC}(i), which clearly shows the presence of only delocalized states at $W/t=1.2$. The $D(2)$ value is calculated as $0.99$ with the system size 
$L=6765$. In Figs.~\ref{Fig:MFS-1p5}(ii)-(iv), we have fixed the disorder strength beyond the critical point at $W/t=3.0$. In these figures we have performed the multifractal analysis of the states corresponding to the different regions, marked as $ERG^1, ERG^2$ and $LOC^1$ in the Fig.\ref{Fig:IPR-SO-a1p5-LOC}(ii). We primarily focus on these three regions since it is sufficient to prove the existence of the multiple mobility edges which separate alternative bands of delocalized and localized states. In Figs.\ref{Fig:MFS-1p5}(ii) and (iii), we have shown the multifractal analysis of the eigenstates present in the regions of $ERG^1$ and $ERG^2$. It is quite evident that as the system size increases to a large value, $f(\alpha)$ is non-zero only for $\alpha=1$, clearly indicating that 
these states are delocalized. The $D(2)$ value is calculated for both the regions and found to be $0.99$ with system size $L=6765$. The multifractal spectrum is most crucial to determine the nature of the states of the $LOC^1$ region. From the IPR results, we expect the eigenstates of the $LOC^1$ to be localized. However, relatively large fluctuations in the IPR values in this region, although they are system size independent, makes it difficult to draw a definite conclusion. In Fig.~\ref{Fig:IPR-SO-a1p5-LOC}(iv) we have presented the multifractal spectrum for these states. In contrast to the other regions, $f(\alpha) \rightarrow 0$ as $\alpha \rightarrow 0$ with an increase in the system size, which shows that these states are indeed localized. We have estimated the $D(2)$ value of $LOC^1$ for the system size $L=6765$, and it is found to be $0.1$. 

In Fig.~\ref{Fig:MFS-0p5}, a similar analysis has been carried out in the long-range $(a \leq 1)$ limit of the GAA Hamiltonian in the presence of
the RSO coupling. The results are presented for two different disorder strengths $W/t= 1.0$ and $W/t= 4.0$, where the critical value $W_c/t \approx 2.1$. Similar to the previous analysis, we have set $\alpha_y=0 $, and $\alpha_z\neq0$ in this case well. From Fig.~.\ref{Fig:IPR-SO-a0p5-MFS}(ii), we have concluded that above the critical point the spectrum breaks into alternative bands of delocalized and multifractal states, a few of which have been labelled as $ERG^1$, $ERG^2$, and $MF^1$. For computing the multifractal spectrum of Fig.~\ref{Fig:MFS-0p5}(i) we have considered all the eigenstates. The disorder strength is set at $W/t= 1.0$. It is obvious that the quasipaticle eigenstates are delocalized. The $D(2)$ value is calculated to be $0.99$ with the system size $L=6765$. In Figs.~\ref{Fig:MFS-0p5}(ii)-(iv), we have presented the multifractal spectrum corresponding to the states $ERG^1$, $ERG^2$, and $MF^1$ respectively. The disorder strength is set at $W/t=4.0$. The multifractal spectrum of Figs.~\ref{Fig:MFS-0p5}(ii) and (iii) clearly indicates that the states $ERG^1$ and $ERG^2$ are delocalized. For both these regions,  the estimated value $D(2)$ turns out to be $0.99$ for $L=6765$. In Fig.\ref{Fig:MFS-0p5}(iv), we have presented the multifractal spectrum of the $MF^1$ region. The difference in the spectrum compared to the delocalized and localized states are quite self evident. In this case, as the system size increases $f(\alpha) \rightarrow 0$ while $\alpha \neq 0$ which clearly demonstrate that these states are of multifractal in nature. As expected, the $D(2)$ value of this $MF^1$ region is found to be $0.48$.
\begin{figure*}[ht]
\centering
\includegraphics[width=0.35\textwidth,height=0.30\textwidth]{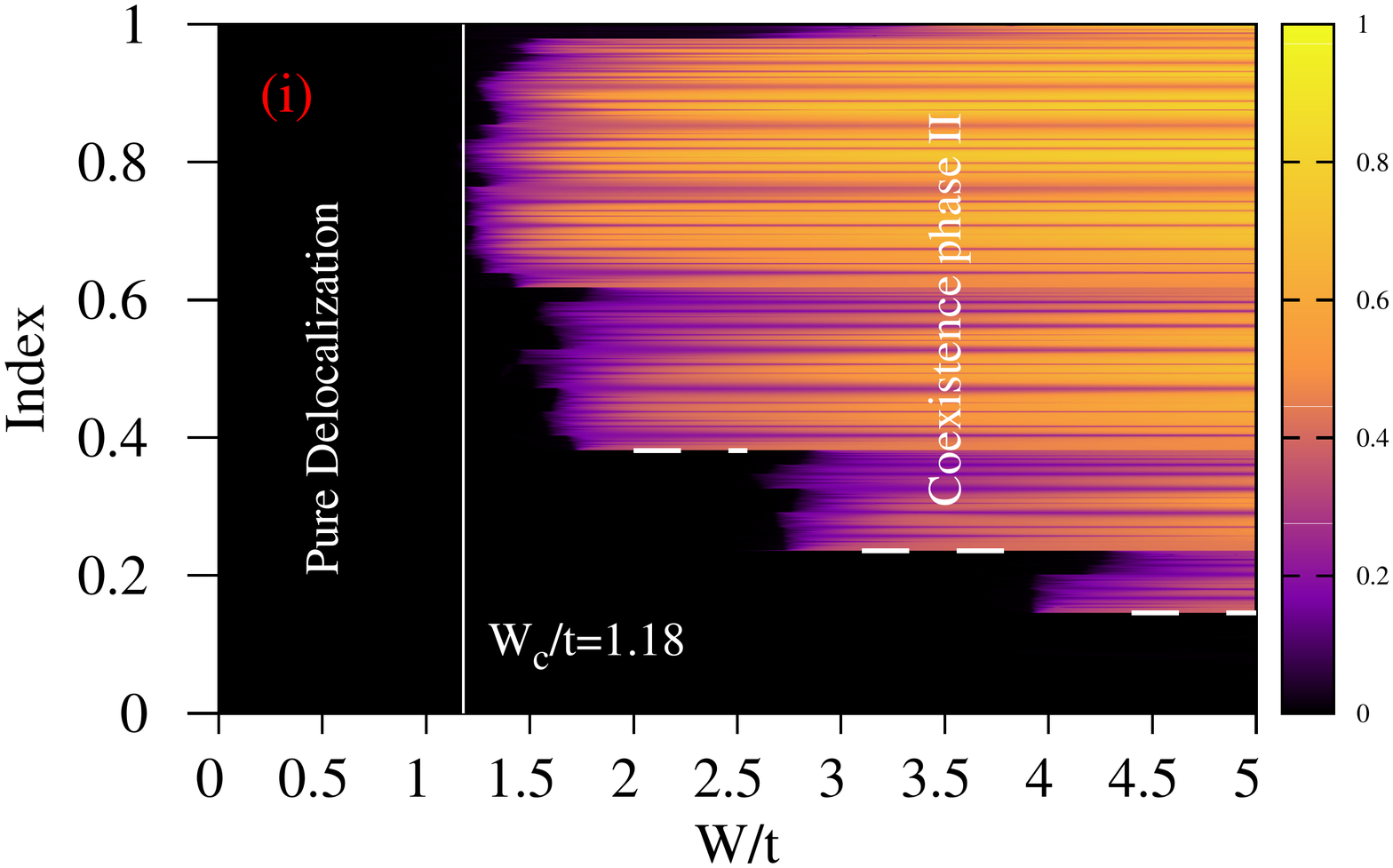}\hspace{-0.8cm}
\includegraphics[width=0.35\textwidth,height=0.30\textwidth]{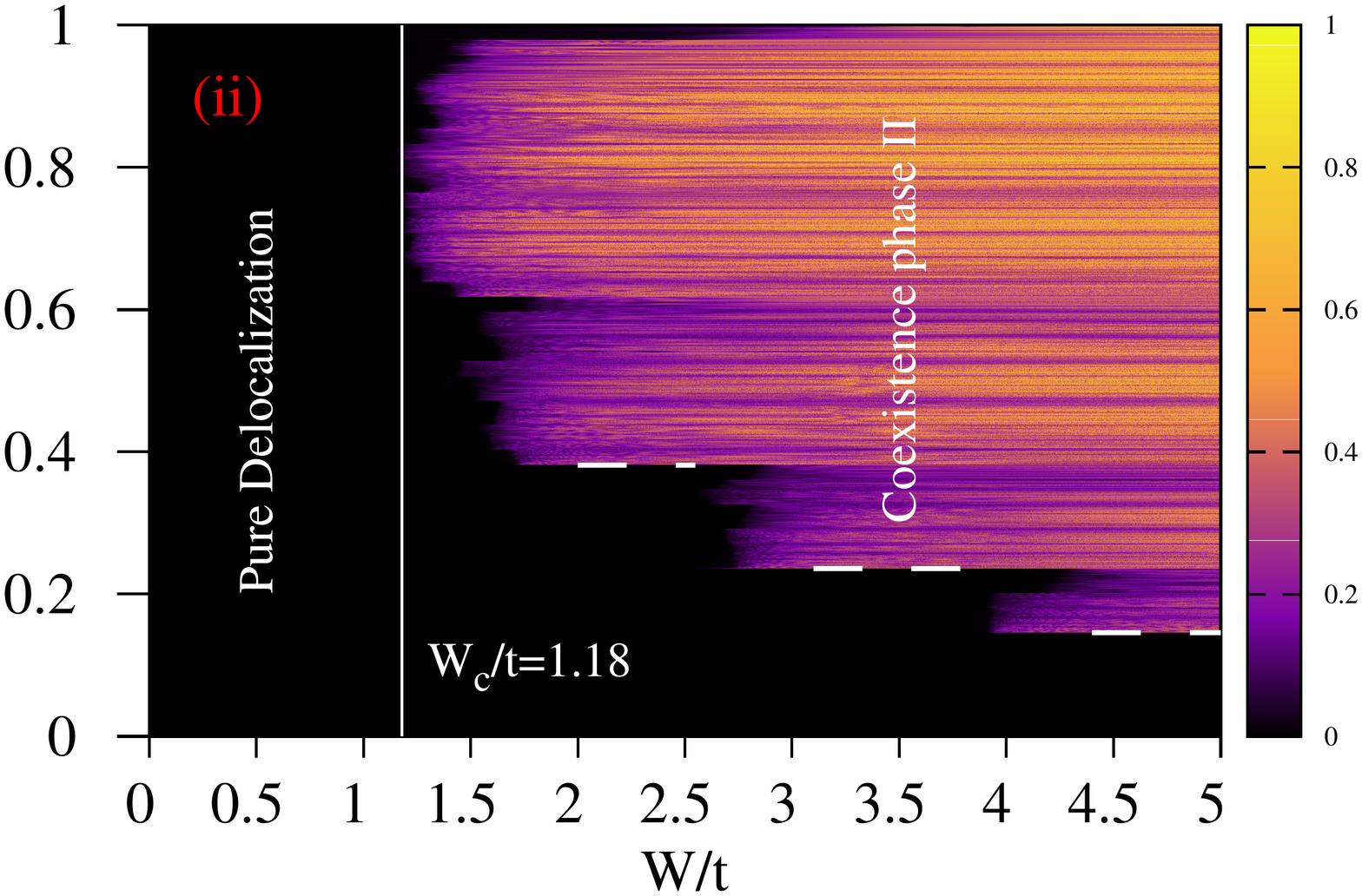}\hspace{-0.8cm}
\includegraphics[width=0.35\textwidth,height=0.30\textwidth]{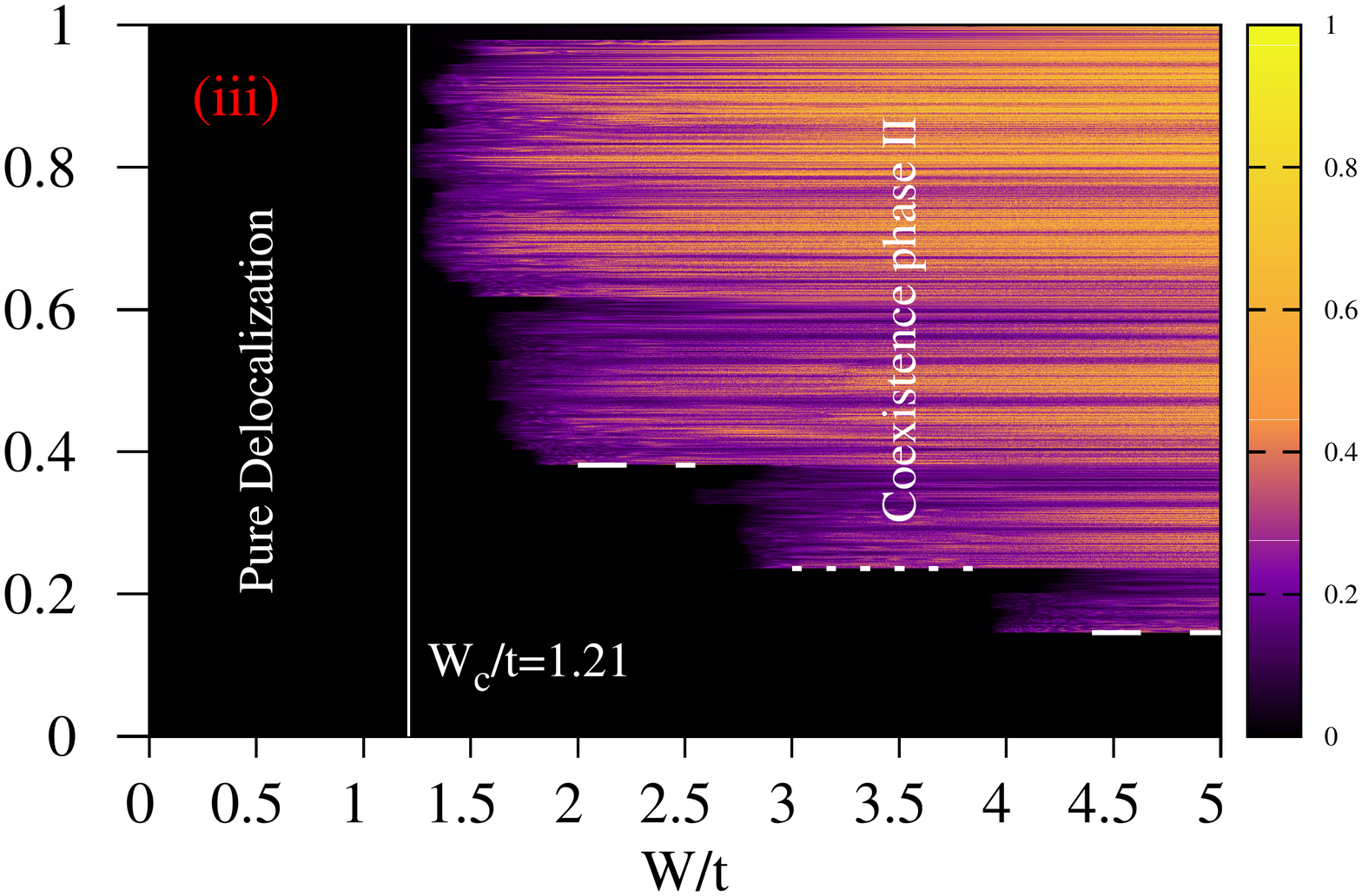}\hspace{-0.8cm}
\caption{Projection of IPR as function of $W/t$ and eigen-energies $(index)$ in the limit of the short-range $(a = 1.5)$ GAA model with RSO coupling . (i) $\alpha_y=0.1,$ $\alpha_z=0.0$, (ii) $\alpha_y=0.0,$ $\alpha_z=0.1$, and (iii) $\alpha_y=0.1,$ $\alpha_z=0.1$. Dotted lines show the presence of single mobilty edge regions.}
\label{Fig:IPR-1p5-small}
\end{figure*}
\section{Conclusions}
In conclusions, we have studied the effect of spin-orbit coupling of Rashba type on the one-dimensional quasiperiodic lattice, described by the GAA Hamiltonian, in the short-range and long-range hopping limits. In the original GAA Hamiltonian, it had been reported earlier that there is a single mobility edge in the short-range hopping limit beyond a critical disorder strength. In contrast, a single multifractal edge separates delocalized and multifractal states. In this work, at first, we have found that while this is true in the long-range hopping limit, in the short-range hopping limit, there exist certain windows of disorder strengths where a single edge separates delocalized and localized states. Outside these windows, in the short-range hopping limit, we have found, in contrast to previously reported results, that a narrow band of delocalized states can exist after the appearance of the first mobility edge. In the original GAA Hamiltonian, in both of these limits, where single mobility (multifractal) states exist, the lowest $\beta^s L$ ($s=1,2,\cdots$) states are delocalized, and all the states above it are localized (multifractal type). This article demonstrates that spin-orbit coupling has a dramatic effect on the energy spectrum, especially in the long-range hopping limit. In general, in the presence of the spin-orbit coupling, the critical disorder strength increases to a larger value compared to the original GAA Hamiltonian in both these limits. In the short-range hopping limit, the windows of disorder strengths where a single mobility edge completely separates delocalized and localized states get successively destroyed with an increase in the spin-orbit coupling strength, and multiple mobility edges can be found outside these windows. The spin-orbit coupling similarly affects the windows mentioned above in the long-range hopping limit. In this case, they also get successively destroyed with an increase in the spin-orbit coupling strength. Furthermore, in the presence of the spin-orbit coupling, multiple multifractal edges appear in the spectrum that separate alternative bands of delocalized and multifractal states beyond the critical point. Interestingly, in contrast to the short-range hopping limit, these bands of the delocalized states survive up to a very large value of disorder strength. We have found that the multifractal edges roughly appear at $n/L = \beta^s \pm \beta^m$, where $s, m= 0,1,2,\cdots$.

\begin{acknowledgements}
The authors would like to acknowledge the computational facility provided by SERB (DST), India (EMR/2015/001227). 
\end{acknowledgements}

\appendix
\section{}\label{Sec:RSO-IPR-Small}
In this section, we have discussed the behavior of mobility edges in the energy spectrum for a short-range $(a>1)$ GAA model having a smaller 
amplitude of RSO coefficients $\alpha_y,\alpha_z$ (as shown in Fig.\ref{Fig:IPR-1p5-small}). We can observe that with weaker RSO coupling, 
only the first window (where only a single mobility edge exists) just after the critical point gets destroyed, while the rest of such windows 
remain unaffected. In Fig.~\ref{Fig:IPR-SO-a1p5} of the main text, we have observed that as the strength of the RSO coupling is increased,
such windows appear only at a higher strength of the disorder. 

\end{document}